\documentclass[journal]{IEEEtran}
\usepackage{graphicx}
\usepackage{overpic}
\usepackage{amsmath}
\usepackage{amssymb}
\usepackage{multirow}
\usepackage{makecell}
\usepackage{array}
\usepackage{stfloats}
\usepackage{gensymb}
\usepackage{xcolor}
\usepackage{float}
\usepackage{booktabs}
\usepackage{cite}
\usepackage{hyperref}
\hypersetup{
	colorlinks=true,
	linkcolor=cyan,
	urlcolor=red,
	citecolor=cyan,
}

\begin{document}

\title{Synthetic Aperture Radar Image Change Detection via Siamese Adaptive Fusion Network}
\author{Yunhao Gao, Feng Gao, Junyu Dong, Qian Du, Heng-Chao Li
\thanks{This work was supported in part by the National Key Research and Development Program of China under Grant 2018AAA0100602, in part by the National Natural Science Foundation of China under Grant U1706218 and Grant 61871335, and in part by the Key Research and Development Program of Shandong Province under Grant 2019GHY112048. ({\it Corresponding author: Feng Gao})}
\thanks{Y. Gao, F. Gao, and J. Dong are with the Qingdao Key Laboratory of Mixed Reality and Virtual Ocean, School of Information Science and Engineering, Ocean University of China, Qingdao 266100, China. (e-mail: gaofeng@ouc.edu.cn)}
\thanks{Qian Du is with the Department of Electrical and Computer Engineering, Mississippi State University, Starkville, MS 39762 USA.
}
\thanks{H.-C. Li is with the Sichuan Provincial Key Laboratory of Information Coding and Transmission, Southwest Jiaotong University, Chengdu 610031, China.} 
}

\markboth{IEEE Journal of Selected Topics in Applied Earth Observations and Remote Sensing}%
{Shell}

\maketitle
\begin{abstract}
\textcolor{blue}{This work has been accepted by IEEE JSTARS for publication.}
Synthetic aperture radar (SAR) image change detection is a critical yet challenging task in the field of remote sensing image analysis. The task is non-trivial due to the following challenges: Firstly, intrinsic speckle noise of SAR images inevitably degrades the neural network because of error gradient accumulation. Furthermore, the correlation among various levels or scales of feature maps is difficult to be achieved through summation or concatenation. Toward this end, we proposed a siamese adaptive fusion network for SAR image change detection. To be more specific, two-branch CNN is utilized to extract high-level semantic features of multitemporal SAR images. Besides, an adaptive fusion module is designed to adaptively combine multiscale responses in convolutional layers. Therefore, the complementary information is exploited, and feature learning in change detection is further improved. Moreover, a correlation layer is designed to further explore the correlation between multitemporal images. Thereafter, robust feature representation is utilized for classification through a fully-connected layer with softmax. Experimental results on four real SAR datasets demonstrate that the proposed method exhibits superior performance against several state-of-the-art methods. Our codes are available at \verb'https://github.com/summitgao/SAR_CD_SAFNet'.

\end{abstract}


\IEEEpeerreviewmaketitle

\section{Introduction}

\IEEEPARstart{R}{mote} sensing image change detection aims to find the changed information between two multitemporal images acquired in the same area at different times \cite{Quin13_tgrs}. It provides valuable information for many applications, such as target detection \cite{Quin10_pr}, natural resource supervision \cite{Bruzzone97_tgrs}, and agricultural development \cite{Radke05_tip}. When a natural disaster suddenly occurs, a robust change detection algorithm can efficiently detect subtle changes, and corresponding measures can be quickly taken by local governments to reduce the loss of life and property \cite{Jian14_tcb} \cite{zhang21_tcb}. Therefore, change detection has attracted extensive research attention.

Synthetic aperture radar (SAR) images are produced by an active system that sends a signal to the ground, and then receives the reflected signal. Different objects exhibit different characteristics between the scattering and polarized signals, which is beneficial to further accurate interpretation \cite{Liu20_tnnls}. The SAR sensor has an all-weather and all-time imaging capability. It can penetrate smoke, cloud, and haze to acquire high-quality images \cite{Li15_grsl}. Especially when a natural disaster occurs, SAR images make up the shortcomings of other data sources, such as LiDAR, optical, and multispectral data. Therefore, SAR imagery is a very useful data source for change detection.

It is worth mentioning that some pioneer efforts have been devoted to tackling the SAR change detection task. These methods can be classified into two broad categories: supervised and unsupervised approaches \cite{Chen19_nc}. The supervised approaches usually achieve better performance than the unsupervised ones by learning from labeled samples. However, it is generally difficult to collect high-quality labeled samples or acquire prior knowledge of the region to be studied. Therefore, the supervised approaches are commonly combined with an unsupervised one. The unsupervised approaches are employed for reliable sample generation, and the tedious task of manually labeling samples is eliminated.

In this paper, we focus on unsupervised SAR image change detection. Due to speckle noise, it is very challenging to identify changes accurately in SAR images. To overcome the challenges, researchers designed a framework with three steps: image preprocessing, difference image (DI) generation, and DI classification \cite{Bruzzone02_tip}. The first step involves denoising and coregistration. Denoising reduces speckle noise to some extent, but it may cause undesired degradation of the geometric details. Coregistration with subpixel accuracy is critical to generate a robust DI. In the second step, the ratio method \cite{Bazi05_tgrs} is commonly used to generate a DI. Many efficient operators are proposed, such as the log-ratio operator \cite{Dekker98_ijrs}, Gauss-ratio operator \cite{Hou14_jstars}, and neighborhood-based ratio operator \cite{Gong12_grsl}. In the final step of change detection, thresholding, maximum expectation, and clustering methods are generally involved for classification. In \cite{Krinidis10_tip}, fuzzy local information c-means (FLICM) clustering algorithm was proposed to provide robustness to noisy images. In \cite{Gong12_tip}, changed and unchanged pixels were clustered by the fuzzy c-means (FCM) based on Markov random field (MRF) energy function. Gao \emph{et al.} \cite{Gao16_grsl} proposed the PCANet to further classify the preclassification results. That achieves excellent performance in speckle noise suppression. In addition, many advanced methods have been employed for change detection, such as extreme learning machine (ELM) \cite{Gao16_jars} and support vector machines (SVM) \cite{Wang16_rsl}. However, it is difficult to capture high-level semantic features for change detection automatically and effectively.

Recently, deep learning-based methods have become increasingly popular and shown their superiority in remote sensing communities \cite{Gao21_tgrs}. Many attempts in remote sensing image analysis based on deep learning methods have been motivated by these successful applications \cite{Gong17_tec, Jian18_jvcir, Li20_tnnls, Wang21_tgrs}. Some researchers make efforts to design deep models for the change detection task. In these models, unsupervised clustering methods are first employed to preclassify the input SAR images, and training samples are selected from reliable labels based upon preclassification. These samples are considered as prior knowledge and fed into a deep model for training \cite{Jian21_esa}. Finally, the model forms its interpretation of the input image, and then the final change map is obtained. Gong \emph{et al.} \cite{Gong15_tnnls} first presented a SAR change detection method based on deep learning. The stacked restricted Boltzmann machines (RBMs) are utilized for feature extraction. Liu \emph{et al.} \cite{Liu16_tnnls} presented a change detection framework based on a convolutional coupling network. The network is symmetric with each side consisting of one convolutional layer and several coupling layers. Some recent breakthroughs in change detection were achieved by the convolutional neural network (CNN). Mou \emph{et al.} \cite{Mou18_tgrs} proposed a change detection framework that combines CNN and recurrent neural network into an end-to-end network. Gao \emph{et al.} \cite{Gao19_grsl} detected changed information from sea ice SAR images by transferred deep learning. In \cite{Wang19_tgrs}, a general end-to-end 2-D CNN named GETNET was designed for hyperspectral image change detection. Liu \emph{et al.} \cite{Liu18_tnnls} presented a local restricted CNN framework for SAR change detection, in which the original CNN is improved with a local spatial constraint. In \cite{Li19_rs}, a noise modeling-based unsupervised fully convolutional network (FCN) framework was presented for HSI change detection, which was proved with powerful learning features. However, a single-branch neural network inevitably results in the accumulation of error gradients. In other words, unstable feature representation limits its performance.

It is non-trivial to build an effective SAR change detection model, due to the following two challenges: 1) \emph{Unstable feature representation}. Intrinsic speckle noise of SAR image inevitably degrades the neural network because of error gradient accumulation. It affects the performance of change detection to some extent, which results in an unstable network. Moreover, it is difficult to make full use of complementary information between different levels of features. Therefore, the extracted features may not well describe the changed information between multitemporal SAR images. 2) \emph{Insufficient feature correlation}. Feature representations from two images are employed to represent the changed information, but the correlation may not be well explored through fusion operations such as summation and concatenation. 

To solve the above challenges, we establish a deep \underline{S}iamese \underline{A}daptive \underline{F}usion \underline{Net}work (SAFNet) for SAR image change detection, which exploits stable feature representation based on siamese architecture. Specifically, we employ two-branch networks to exploit the high-level semantic features, which are rarely considered in SAR image change detection tasks. Furthermore, a correlation layer is designed to integrate features from two-branch networks. The attention mechanism is introduced to adaptively choose features among different scales. Ultimately, we introduce conditionally parameterized convolutions to enhance feature representation. Extensive experiments on four real SAR datasets demonstrate the superiority of the proposed SAFNet over state-of-the-art methods. Meanwhile, we have released our codes to facilitate other researchers.

The main contributions of the proposed SAFNet are summarized as follows:

\begin{itemize}
	
	\item We explore the SAR change detection task via a well-designed siamese neural network. The two branch network independently extracts the features of multitemporal images, and then the discrimination of features is improved through similarity measurement.
	
	\item For stable feature representation, an adaptive fusion module is utilized to combine the outputs of different layers. Since features from different layers contain complementary information, the attention-based fusion mechanism is introduced to use such information to improve feature representation. Therefore, the multiscale responses from convolutional layers are adaptively fused. 
	
	\item To avoid the loss of correlation caused by traditional feature integration methods, such as summation and concatenation, a correlation layer is designed for feature integration.
	
\end{itemize}

The rest of this paper is organized as follows. In Section II, the proposed method is described in detail. Section III presents the experimental results on real multitemporal SAR images to validate the proposed method. Finally, the conclusion is drawn in Section IV with plausible future works.

\section{Methodology}

Given two coregistered SAR images $I_1$ and $I_2$ captured at different periods, we aim to generate a change map that represents the changed information between two images. The proposed change detection method is comprised of two steps: Firstly, high-level semantic features of multitemporal images are extracted through two-branch networks, and a similarity measure is used to optimize the process of feature extraction. Secondly, a correlation layer is employed to integrate the features for classification and change map generation.

\begin{figure*}[t!]
	\centering
	\includegraphics[width=5.8in]{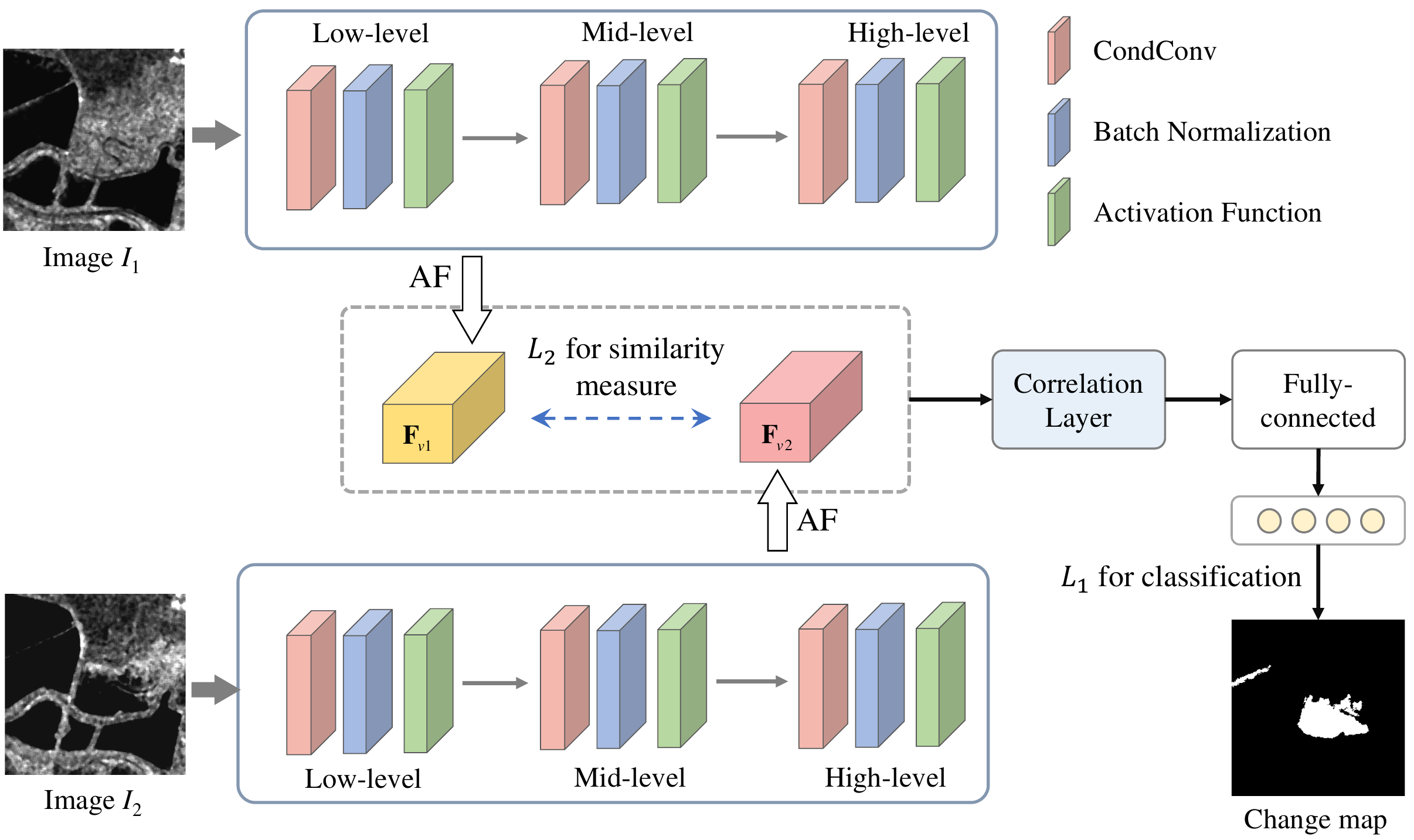}
	\caption{Framework of the proposed Siamese Adaptive Fusion Network (SAFNet). Low-level, mid-level, and high-level CondConv blocks are utilized to extract three levels of features. Then fuse them by adaptive fusion (AF) module. Finally, the final label prediction is achieved by the correlation layer and fully-connected layer.} 
	\label{fig_safnet}
\end{figure*}

\subsection{Feature Extraction and Reliable Sample Generation}

In this work, spatial neighborhood information of each pixel is analyzed for label prediction, thus detecting the changed region (the pixel with label ``1") in the image. Given two multitemporal SAR images $I_1$ and $I_2$, we first obtain the pixel-wised patch-pairs $(x_i^{t_1},x_i^{t_2},y_i)$, where $x_i^{t_1} \in \mathbb{R}^{r\times r}$ and $x_i^{t_2} \in \mathbb{R}^{r\times r}$ are the image patches centered the $i$-th pixel in $I_1$, and $y_i$ is the ground truth label. Therefore, the spatial information of the $i$-th pixel is fed into the CNN for feature extraction. 

The traditional change detection model usually analyzes the difference image generated by log-ratio or neighborhood ratio. Therefore, change detection results depend heavily on the quality of the difference image. Especially, the performance always deteriorates when SAR images are seriously disturbed by speckle noise. In this paper, feature representation is directly extracted from the original SAR images by a two-branch CNN model, which is less sensitive to speckle noise. Therefore, stable feature representation can be achieved. The proposed SAFNet is comprised of two branches $(S_1, S_2)$ that accept patch-pairs as input. The feature extraction step is expressed as:
\begin{equation}
\mathcal{F}_{merge}^i= \mathcal{C}(\mathcal{F}_1^i, \mathcal{F}_2^i)) =\mathcal{C}( S_1(x_i^{t_1}), S_2(x_i^{t_2})),
\end{equation}
where $\mathcal{F}_1^i$ denote the features of image patch $x_i^{t_1}$ extracted by $S_1$, $\mathcal{F}_2^i$ denote the features of image patch $x_i^{t_2}$ extracted by $S_2$, and  $\mathcal{C(\cdot)}$ denotes the correlation analysis. It is implemented by a group convolution operation, where $\mathcal{F}_1^i$ and $\mathcal{F}_2^i$ denote the input and kernels respectively. Then, the final fused features $\mathcal{F}_{merge}^i$ are obtained for classification by a fully-connected layer with softmax. In the process of network training, similarity measure and classification loss are applied to optimize the SAFNet. The pseudo-label samples are generated by the FCM algorithm in an unsupervised manner \cite{Gao16_jars}. We randomly select a certain proportion of samples from the pseudo label set for training, and the rest for testing. The details of pseudo-label sample generation are described as follows:

\begin{itemize}
	\item [1:]The FCM algorithm is performed on DI to generate the changed and unchanged clusters: $\Omega_c^1$ and $\Omega_u^1$. Here, the number of changed clusters is $T_c^1$. The upper limit of the change class is no more than $T_c^1\cdot \theta$, $\theta = 1.2$.
	\item [2:]The FCM algorithm is reperformed on DI to generate five clusters: $\Omega_1^2$, $\Omega_2^2$, $\Omega_3^2$, $\Omega_4^2$, $\Omega_5^2$. $\Omega_1^2$ has the higher probability of changed clusters, and so on. The number of clusters is $T_1^2$, $T_2^2$, $T_3^2$, $T_4^2$, $T_5^2$. The pixels from $\Omega_1^2$ is assigned to $\Omega_c$.
	\item [3:] The clusters are assigned to $\Omega_i$ when $\sum\limits_{i=1} T_i^2<T_c^1\cdot \theta$, and the rest clusters are denoted as $\Omega_u$. Thus, the preclassification map consisted of $[\Omega_c, \Omega_i, \Omega_u]$ is generated.
		
\end{itemize}

\subsection{Siamese Adaptive Fusion Network}

In this paper, we propose SAFNet to compare image patches from SAR images, and it achieves robust feature discrimination power through adaptive fusion (AF) module and correlation layer. The framework of the proposed SAFNet is illustrated in Fig. \ref{fig_safnet}.

Image patches centered at selected sample pixels are extracted from $I_1$ and $I_2$, respectively. Two groups of patch-pairs are treated as the input of two branches. The network optimization in SAFNet is achieved by weight sharing and residual learning. Each branch of the SAFNet is comprised of three CondConv blocks and the AF module, which are merged by the correlation layer and fully-connected layer. In the following, we will describe the SAFNet in detail.

\subsubsection{The CondConv Blocks}

\begin{figure}[h]
	\centering
	\includegraphics[width=3.4in]{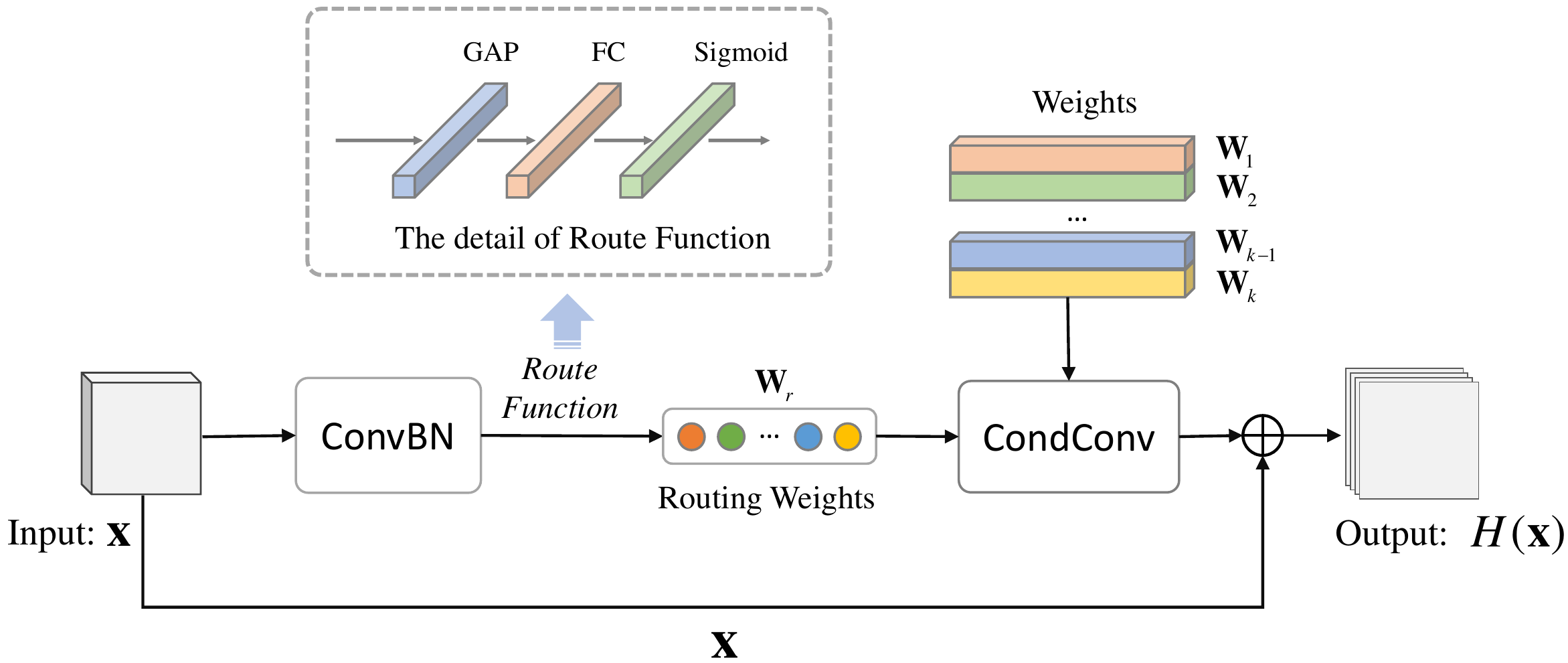}
	\caption{Detailed components of the CondConv block. ConvBN denotes three operations: convolution, batch normalization and activation layer. The conditionally parameterized convolutions (CondConv) are achieved by combining initialization weight and routing weight from route function. Therefore, the ability of feature extraction is significant with one kernel computation.}
	\label{fig_condconv}
\end{figure}

Although feature extraction is greatly improved by aggregating multiple convolutional branches, the computational cost increases dramatically \cite{Xie17_cvpr}. To address the problem, conditional computation was performed by activating only a portion of the entire network \cite{Bengio13_arxiv}. In \cite{Yang19_nips}, efficient inference was performed by conditionally parameterized convolutions.

The structure of a typical conditionally parameterized convolutions (CondConv) block is shown in Fig. \ref{fig_condconv}, which generates several groups of convolution kernels through routing function and initialization weights, and the details of CondConv are described in \cite{Yang19_nips}. ConvBN is the combination of three operations: convolution, batch normalization, and activation layer. 

The routing weights $\textbf{W}_r$ generated by routing function is calculated as:\begin{equation}
\textbf{W}_r = R(\textbf{F}_{c1})=\delta(GAP(\textbf{F}_{c1})*\textbf{W}_f),
\end{equation}
where $GAP$ denotes the global average pooling, $\delta$ is the Sigmoid function, and $*$ is the convolution operation. $\textbf{W}_f$ denotes the matrix of fully-connected layer, which mapping the global feature to $k$ routing weights. Therefore, the output of CondConv through residual learning \cite{He16_cvpr} is defined as:
\begin{equation}
H(x) = \sigma((\alpha_1 \cdot \textbf{W}_1+\cdots + \alpha_k \cdot \textbf{W}_k)*\textbf{F}_{c1}) + \textbf{x},
\label{condconv}
\end{equation}
where $[\alpha_1, \cdots, \alpha_k]$ are the weights from the routing weights $\textbf{W}_r$. Consequently, the CondConv is mathematically equivalent to a linear mixture of multiple convolutions.

\subsubsection{Feature Fusion by Adaptive Fusion Module}

\begin{figure*}[t]
	\begin{center}
		\includegraphics [width=5.5in]{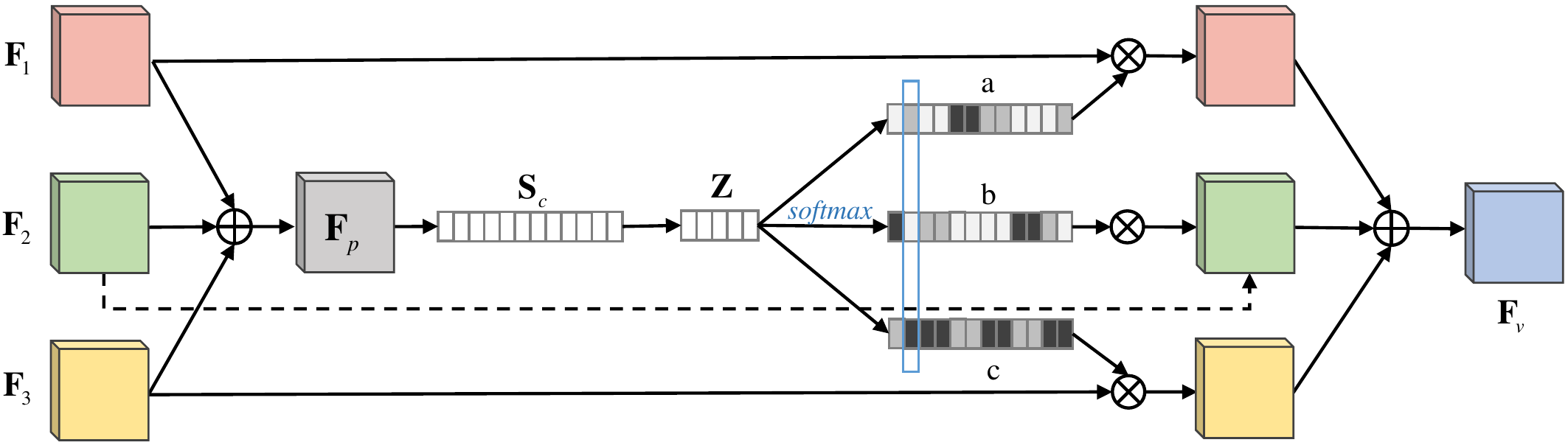}
		\caption{Detail components of the adaptive fusion module.}
		\label{af}
	\end{center}
\end{figure*}

Feature fusion exhibits good performance in many image classification and object detection tasks. Feature fusion is considered as an effective method for complementarity, however, the redundant features are introduced which degrade the discrimination. In the proposed SAFNet, an adaptive feature fusion mechanism is introduced to extract the complementary information among different CondConv blocks. As illustrated in Fig. \ref{fig_safnet}, multiple CondConv blocks are used to capture features of different levels, including low-level, mid-level, and high-level features. 

The important part of each layer of features is given higher weight through a non mutually exclusive attention vector, so the discrimination of the final fused features is stronger.

The output features of the three levels of CondConv blocks (i.e., CondConvBlock1, CondConvBlock2, and CondConvBlock3 in Table \ref{table_safnet}) are denoted as $\textbf{F}_1$, $\textbf{F}_2$ and $\textbf{F}_3$, respectively. One CondConv block is employed to extract the features at each level. In this paper, $\textbf{F}_1$ contains 16 feature maps, $\textbf{F}_2$ contains 32 feature maps, and $\textbf{F}_3$ contains 64 feature maps. Since $\textbf{F}_1$, $\textbf{F}_2$, and $\textbf{F}_3$ contain different number of feature maps, dimension matching is essential for feature fusion. To achieve this, 64 kernels of size $1\times 1$ with different strides are employed to convolute $\textbf{F}_1$, $\textbf{F}_2$, and $\textbf{F}_3$. After such convolution, the feature maps from three levels all become 64 dimensions with the same spatial size for fusion. 

Recently, computer vision adaptively encodes informative context from a long-range region to select more critical information for the current task. Hu \emph{et al.} \cite{Hu18_cvpr} proposed the squeeze-and-excitation network (SENet) to explore the channel relationship of features. In \cite{Li19_cvpr}, selective kernel network (SKNet) adaptively selects the kernel size through attention mechanism. Inspired by these methods, the adaptive fusion (AF) module is designed to merge feature maps from different levels, where an attention mechanism is introduced to emphasize important features while suppressing unnecessary ones. As illustrated in Fig. \ref{af}, three feature maps $\textbf{F}_1$, $\textbf{F}_2$, and $\textbf{F}_3$ are obtained from multiple CondConv blocks. Recognizing that not all the features are essential for the final classification, the attention mechanism is introduced to adaptively choose features from suitable scales. Firstly, the input feature maps are fused by element-wise summation after dimension matching as:
\begin{equation}
\textbf{F}=D(\textbf{F}_1)+D(\textbf{F}_2)+
D(\textbf{F}_3),
\end{equation}
where $\textbf{F} \in \mathbb{R}^{w \times w \times c}$ is the fused feature. $D(\cdot)$ is the function of dimension matching. Then global average pooling (GAP) is employed to capture the global information of $\textbf{F}_s$ as follows:

\begin{equation}
\textbf{F}_s=\frac{1}{w \times w}
\sum_{i=1}^{w}\sum_{j=1}^{w}\textbf{F}(i,j),
\end{equation}
Then, information is aggregated in $\textbf{F}_s\in\mathbb{R}^{c\times 1}$. After that, $\textbf{F}_s$ is further squeezed into a compact feature $\textbf{F}_z \in \mathbb{R}^{\frac{c}{\gamma}\times1}$ by fully-connected layer as:
\begin{equation}
\textbf{F}_z = \sigma(\textbf{W}*\textbf{F}_s),
\end{equation}
where $\sigma$ is the ReLU activation, and $\textbf{W}\in \mathbb{R}^{\frac{c}{\gamma}\times c}$ is the weighting matrix ( $\gamma = 8$ in this paper). Therefore, the model complexity is reduced significantly. 

Fully-connected layer and soft attention (softmax layer) are used to adaptively select features from suitable scales, which is guided by the compact feature descriptor $\textbf{F}_z$. Let \textbf{a}, \textbf{b}, \textbf{c} $ \in\mathbb{R}^{c\times1}$ represent the soft attention vector obtained by the softmax layer. Note that $\textbf{a}_i$ is the $i$-th element of \textbf{a}, likewise $\textbf{b}_i$ and $\textbf{c}_i$. We can have $\textbf{a}_i + \textbf{b}_i + \textbf{c}_i = 1$ owing to the intrinsic feature of the softmax layer. Finally, the feature map $\textbf{F}_v$ is obtained through the attention weights on various scales:
\begin{equation}
\textbf{F}_v= \textbf{a}\cdot \textbf{F}_{1} +
\textbf{b} \cdot \textbf{F}_{2} +
\textbf{c} \cdot \textbf{F}_{3},
\end{equation}
$\textbf{F}_v$ is utilized to generate the final features of one branch of the SAFNet which is denoted by $feat$.

\subsubsection{Correlation Layer}

\begin{figure}[h]
	\centering
	\includegraphics[width=2.2in]{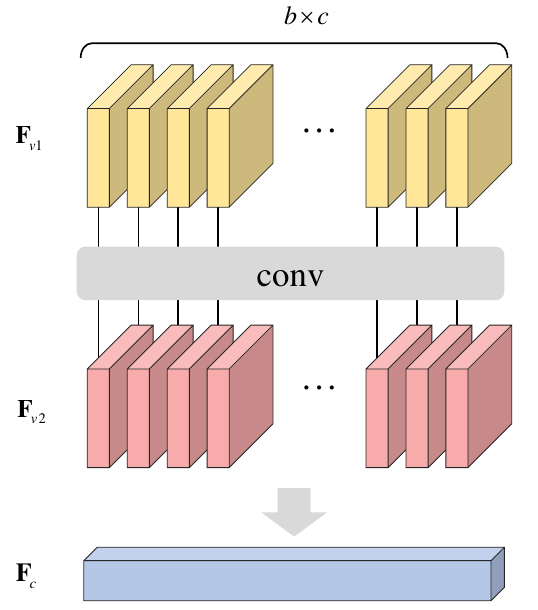}
	\caption{Detailed components of the correlation layer.}
	\label{fig_corre}
\end{figure}

The two branches of networks are employed to extract the semantic features from multitemporal SAR images, and then changed information prediction is achieved after feature integration. Generally speaking, it is easy to ignore the correlation through simple integration operations such as summation and concatenation. Therefore, a feature correlation operation is developed for feature integration. The features from AF modules are denoted as $\textbf{F}_{v1}$ and $\textbf{F}_{v2}$, which can be further integrated through the correlation layer as shown in Fig. \ref{fig_corre}. The integrated features are computed by:
\begin{equation}
\textbf{F}_c = \textbf{F}_{v1} * \textbf{F}_{v2},
\end{equation}
where $*$ denotes the convolution operation (conv), and $\textbf{F}_c \in \mathbb{R}^{b\times1 \times 1 \times c}$ is the final feature representation. It should be noted that the group convolution strategy is applied to the correlation layer. In other words, feature integration through convolution participates in the parameter optimization under the end-to-end training. For the correlation layer, $\textbf{F}_{v1}\in \mathbb{R}^{b\times  w\times w \times c}$ and $\textbf{F}_{v2} \in \mathbb{R}^{b\times w \times w \times c}$ are first reshaped to $\textbf{F}_{v1}\in \mathbb{R}^{1\times k\times  w\times w}$ and $\textbf{F}_{v2} \in \mathbb{R}^{k \times 1\times w \times w}$, respectively. Here, $b$ is the batch size, and $k= b\times c$ denotes the number of kernels. It should be noted that the group convolution strategy is also applied, and the number of groups is $k$.

\subsection{SAFNet Optimization and Change Map Generation}

Table \ref{table_safnet} shows the implementation details of the proposed SAFNet (a typical branch). The input image patch is resized to $28\times28$ pixels. In each branch of the network, three levels of CondConv blocks are implemented. Conv1 and Conv2 are the transitional convolution operators among different levels of CondConv blocks. Each CondConv block contains two convolutions, as illustrated in Fig. \ref{fig_condconv}. After feature extraction, the AF module is employed to emphasize meaningful features. As mentioned before, the SAFNet contains two branches of neural network. Therefore, the SAFNet generates a feature pair $\textbf{F}_{v1}$ and $\textbf{F}_{v2}$. $feat_0$ and $feat_1$ are computed as:

\begin{equation}
feat_0 = \textbf{F}_{v1} * \textbf{W}_{v1}
\end{equation}
\begin{equation}
feat_1 =  \textbf{F}_{v2} * \textbf{W}_{v2}.
\end{equation}

\renewcommand\arraystretch{1.8}
\begin{table*}[h]
	\centering 
	\caption{Details of the proposed SAFNet.}
	\begin{tabular}{ c | c c c c c c}
		\hline 
		Layer & Type & Kernel Size & Stride & Padding & Output & Output Size\\ \hline
		CondConvBlock1 (Low-level)     &Convolution    &3$\times$3   &1  & Yes &16 & 28$\times$28$\times$16 \\   
		Conv1 (Low-level)         &Convolution    &1$\times$1   &2  & No  &32 &14$\times$14$\times$32\\ 
		CondConvBlock2 (Mid-level)    &Convolution    &3$\times$3   &1  &Yes &32 &14$\times$14$\times$32\\ 
		Conv2         &Convolution    &1$\times$1   &2  &No  &64 &7$\times$7$\times$64\\ 
		CondConvBlock3 (High-level)    &Convolution    &3$\times$3   &1  &Yes &64 &7$\times$7$\times$64\\ 
		Correlation layer       & Convolution        &-    &- &-  &64  &1$\times$1$\times$64 \\
		Fully Connected & FC     &-    &- &-  &2 &-\\
		\hline
	\end{tabular}
	\label{table_safnet}
\end{table*}

After obtaining $feat_0$ and $feat_1$, the Euclidean distance metric is employed to produce a value that is used to measure the feature similarity of two branches. To customize an appropriate metric where patch-pairs have stronger discrimination, the contrastive loss is employed. Contrastive loss is formulated as:

\begin{equation}
L_2=\left\{
\begin{aligned}
D(feat_0, feat_1)^2 &  ~~~~~y = 0  \\
\max(0, ~ m-D(feat_0-feat_1))^2 & ~~~~~y=1
\end{aligned}
\right.
\end{equation}
where $feat_0$ and $feat_1$ denote the feature maps extracted from two branches of convolutional network, respectively. $m$ is a margin, which is set as $1$ in this paper, and $D(feat_0-feat_1)$ measures the distance between $feat_0$ and $feat_1$ using Euclidean distance. In this paper, $y$ is the ground truth label that measures the similarity of image pairs. $y=0$ represents higher similarity, and there is no change in the patch-pairs. $y=1$ indicates that the land cover is changed.

To obtain the final classification label in the change detection task, the cross-entropy loss is utilized to optimize the network. The features obtained by the AF module are integrated by correlation operation, and label prediction is achieved through the fully-connected layer and softmax layer. Therefore, the loss value between label prediction and the ground truth is formulated as:

\begin{equation}
L_1= \sum^N_{i=1}[y_t log(y)+(1-y_t)log(1-y)],
\end{equation}
where $y_t$ is the ground truth while $y$ is the label mapping from the last fully-connected layer. $L_1$ is designed to supervise the learning process of the fused feature between multi-temporal SAR images. The ﬁnal loss value is represented as the combination of $L_1$ and $L_2$:

\begin{equation}
L= L_1 + \lambda L_2,
\end{equation}
where $\lambda$ is the weight parameters for the contrastive loss $L_2$. In the experiments, $\lambda$ is empirically set to $0.5$  

Image patch pairs centered at selected samples are used for parameter optimization of the SAFNet. The same as most CNN models, $L$ is optimized using the backpropagation algorithm. Note that $L_2$ can also be considered as regularization terms for $L_1$ to standardize the training process. Finally, the trained SAFNet is utilized for prediction. Each test sample is assigned a label according to the results of SAFNet via a feed-forward propagation (``0" denotes unchanged and ``1" denotes changed ). Then, the final change map is obtained.

\section{Experimental Results and Analysis}

In this section, experiment data and evaluation criteria are introduced firstly. Then we will investigate the factors that may influence the performance of SAFNet. Finally, the effectiveness of SAFNet will be evaluated by comparison with several state-of-the-art methods.

\subsection{Experiment Data and Evaluation Criteria}

\begin{figure}[h]
	\begin{center}
		\includegraphics [width=3.4in]{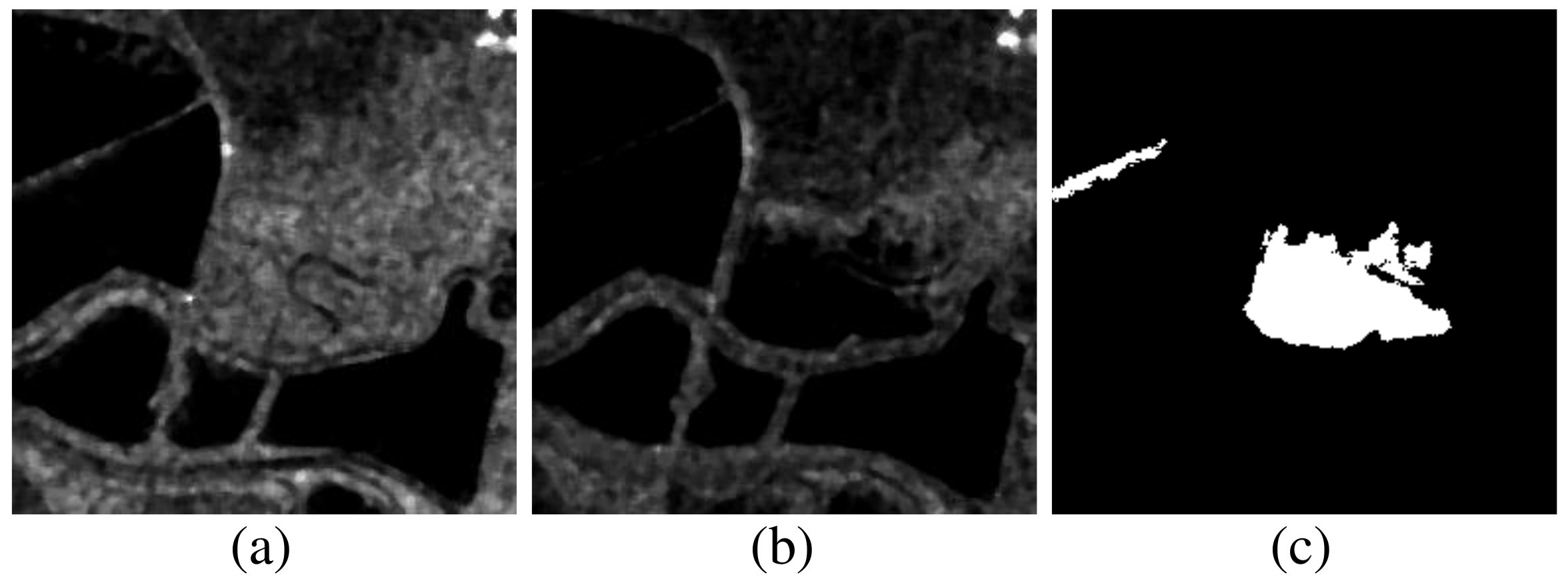}
		\caption{The San Francisco dataset. (a) Image acquired in August 2003. (b) Image acquired in May 2004. (c) Ground truth image.}
		\label{data1}
	\end{center}
\end{figure}

To verify the effectiveness of SAFNet for SAR image change detection, experiments are conducted on four real SAR datasets. The first dataset is the San Francisco dataset. It was acquired by the ERS-2 SAR sensor (the spatial resolution is $30m$) over the city of San Francisco. The size of the original image is $7749\times 7713$ pixels. We select a typical region ($256\times 256$ pixels) to evaluate the proposed method. The images were captured in August 2003 and May 2004, respectively. The ground truth image was created by integrating prior information with photo interpretation. The dataset is shown in Fig. \ref{data1}.

\begin{figure}[h]
	\begin{center}
		\includegraphics [width=3.4in]{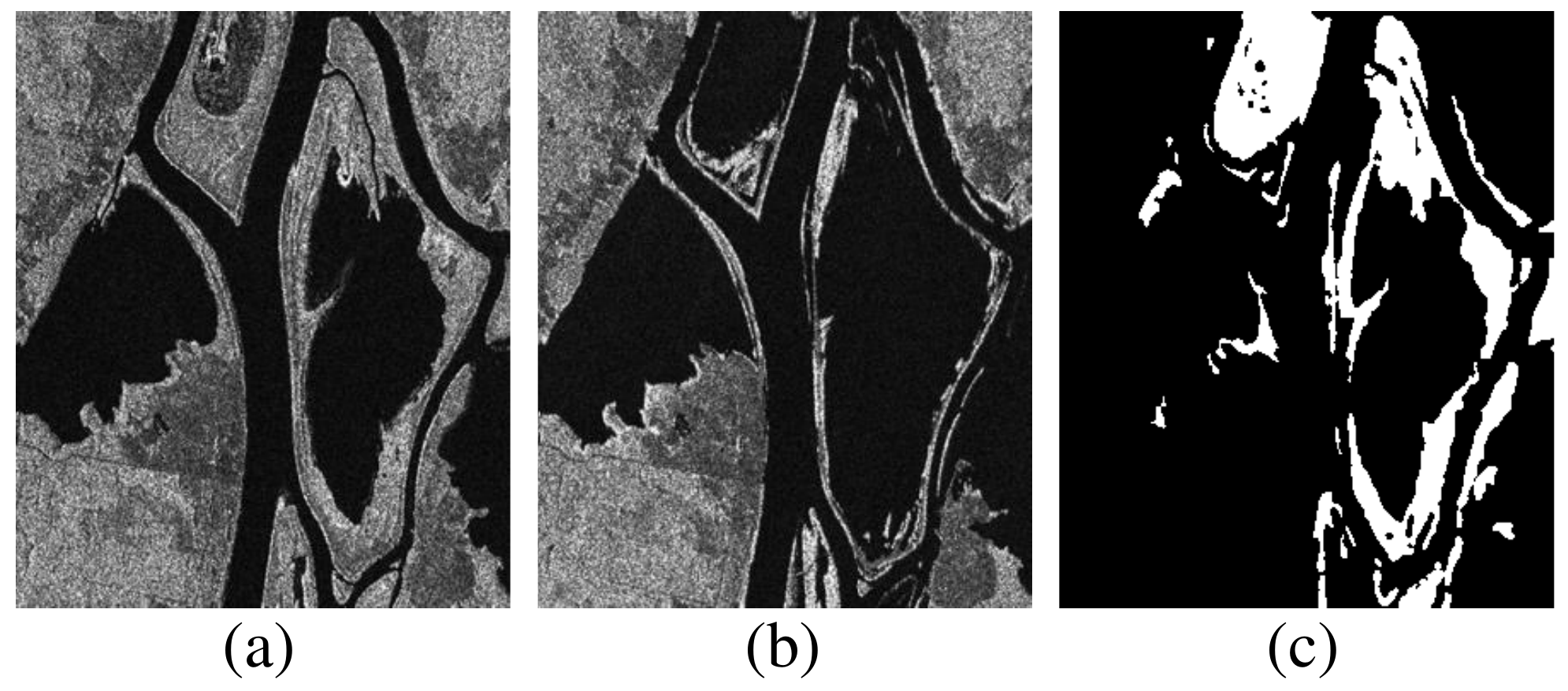}
		\caption{The Ottawa dataset. (a) Image acquired in May 1997. (b) Image acquired in August 1997. (c) Ground truth image.}
		\label{data2}
	\end{center}
\end{figure}

The second dataset is the Ottawa dataset, as shown in Fig. \ref{data2}. Two images were acquired by the Radarsat sensor in May 1997 and August 1997, respectively. The National Defense Research and Development Canada provides the dataset, and the dataset shows the changed information in areas affected by floods. These images were registered by the automatic registration algorithm from A.U.G. Signals Ltd. that is available through distributed computing at {\it{www.signalfusion.com}}. The size of Ottawa dataset is $290\times 350$ pixels and the spatial resolution is $10m$. The available ground truth was created by integrating prior information and photo interpretation.

\begin{figure}[h]
	\begin{center}
		\includegraphics [width=3.4in]{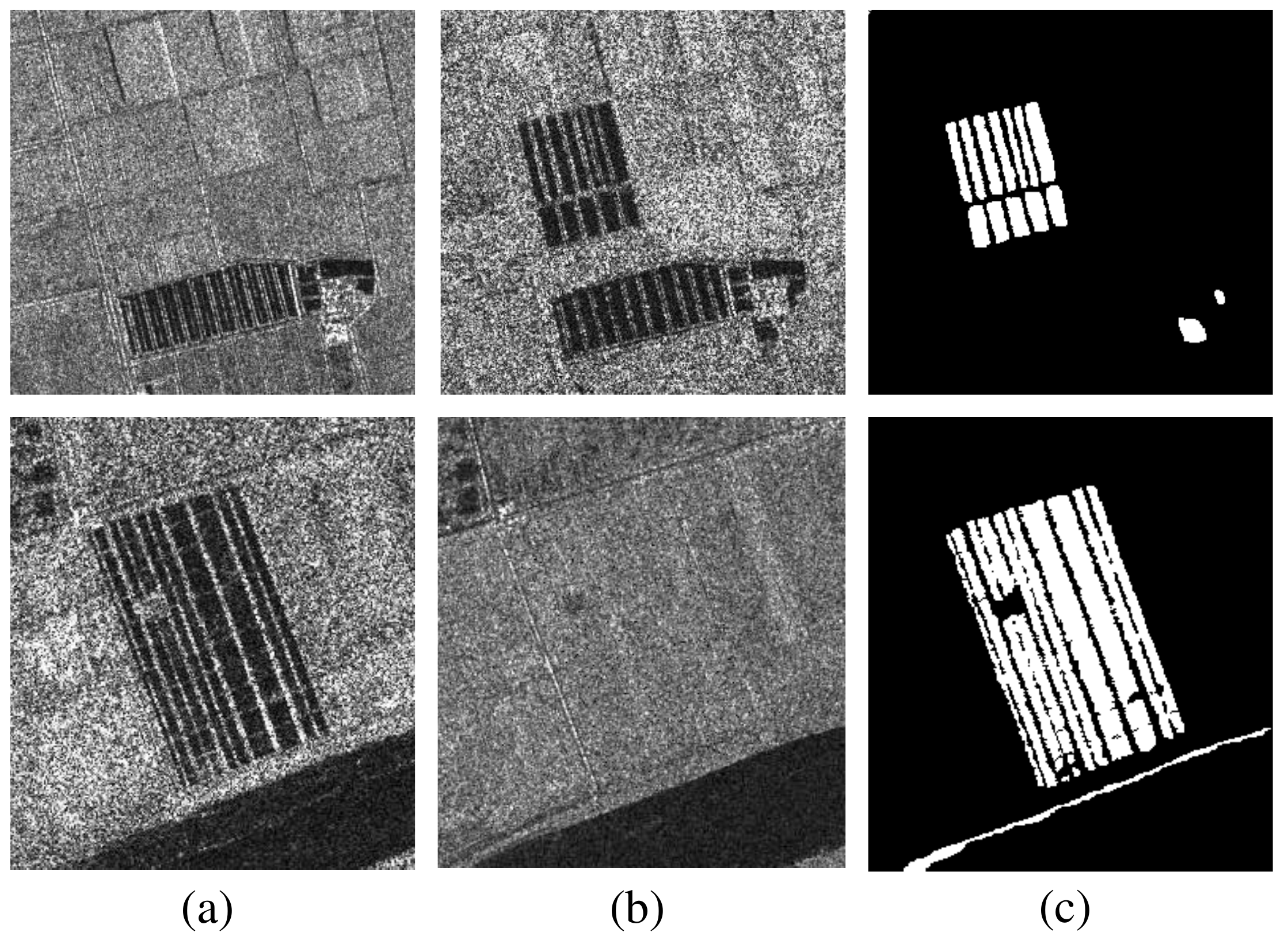}
		\caption{The Yellow River dataset. (a) Image acquired in June 2008. (b) Image acquired in June 2009. (c) Ground truth image.}
		\label{data3}
	\end{center}
\end{figure}

The third dataset is the Yellow River dataset. It was selected from two large SAR images collected in the Yellow River Estuary area of China. Both images were taken by Radarsat-2 (the spatial resolution is $8m$) in June 2008 and June 2009, respectively. The original size of the dataset is $7666\times7692$ pixels. The size is too large to describe details. Therefore, two typical regions ($306\times 291$ pixels and $289 \times 257$ pixels) were used in the experiment to verify the effectiveness of the proposed SAFNet. The changed regions showed newly cultivated farmland. The dataset is shown in Fig. \ref{data3}. In particular, this dataset is polluted by noise with different characteristics. Specifically, one image in the dataset is a single-look image while the other is a four-look image. Therefore, it is very challenging to perform change detection on this dataset.

Both quantitative measures and qualitative analysis are performed on the four datasets. Specifically, the qualitative analysis is achieved by visual comparison between the change map generated by the proposed method and the ground truth image. Besides, false positives (FP), false negatives (FN), overall errors (OE), percentage correct classification (PCC), and Kappa coefficient (KC) are utilized as the quantitative measures to assess the change detection performance. FP denotes the number of unchanged pixels that are mistakenly detected as the changed ones. FN is the number of pixels belonging to the changed class but is incorrectly detected as the unchanged ones. OE denotes the total number of pixels that are incorrectly detected, i.e., the sum of FP and FN. The PCC is computed by:

\begin{equation}
PCC = \frac{N_u+N_c-OE}{N_u+N_c}\times 100\%,
\end{equation}
where $N_u$ denotes the total number of unchanged pixels in the ground truth image while $N_c$ denotes the total number of changed pixels. The KC is computed by:

\begin{equation}
KC = \frac{PCC-PRE}{1-PRE}\times 100\%,
\label{KC1}
\end{equation}
where
\begin{equation}
PRE = \frac{(N_c+FP-FN)\times N_c+(N_u+FN-FP)\times N_u}{(N_u+N_c)\times(N_u+N_c)}.
\label{KC2}
\end{equation}

Here, we can observe that the value of KC is determined by FP and FN rather than by OE alone. Therefore, KC reflects the balance between FP and FN to some extent. More detailed information should be taken into account to obtain suitable KC values, and KC is a more persuasive indicator of the change detection result than PCC or OE.

\begin{figure}[ht]
	\begin{center}
		\includegraphics [width=3.4in]{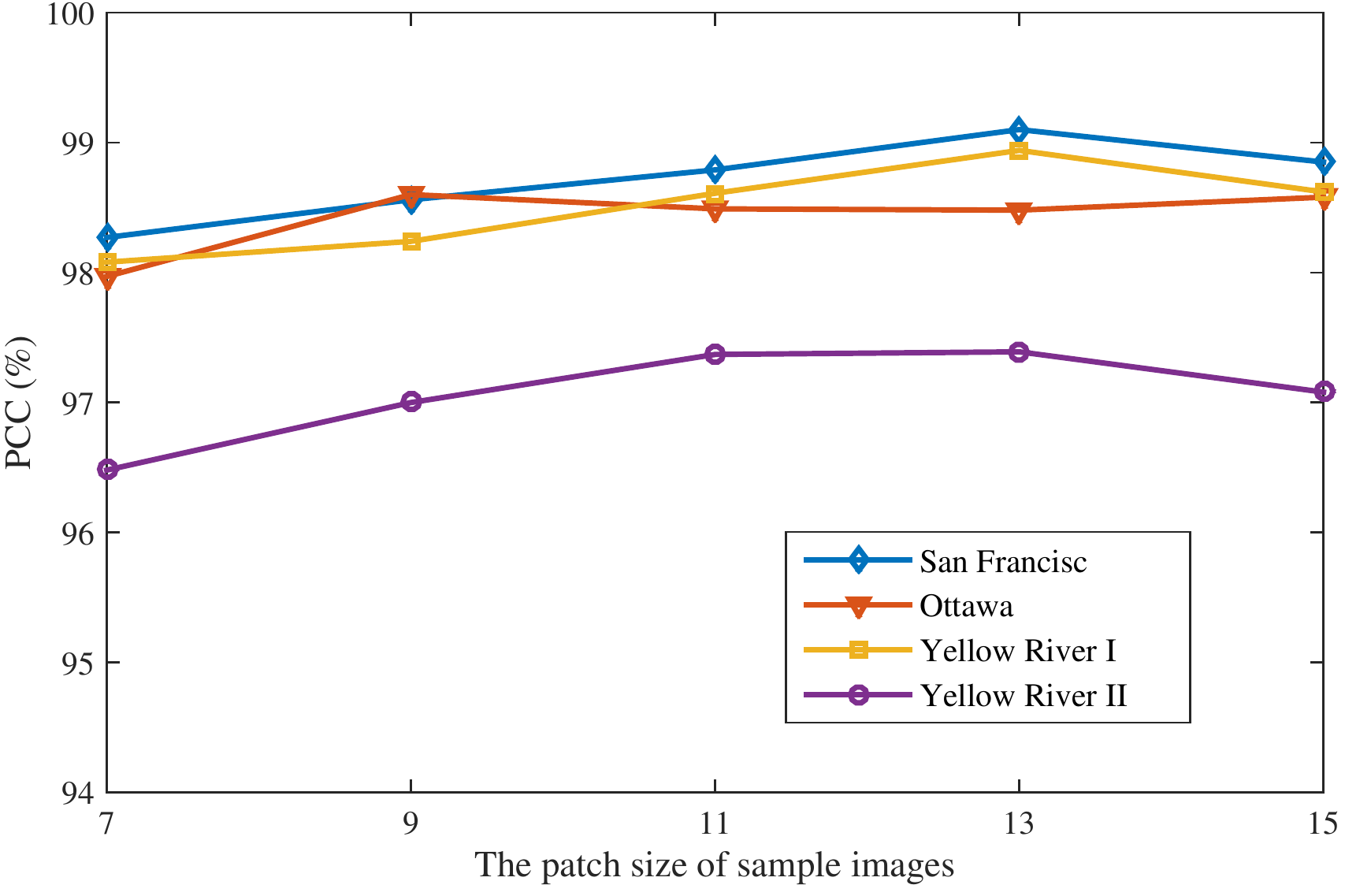}
		\caption{Relationship between patch size $r$ and PCC values.}
		\label{para1}
	\end{center}
\end{figure}

\subsection{Parameter Analysis}

\subsubsection{Analysis of the Sample Image Size}

We first investigate the change detection performance by tuning the values of $r$, which denotes the size of the input patch-pairs employed in SAFNet. We set $r$ to, 7, 9, 11, 13, and 15 to indicate the relationship between $r$ and PCC on four datasets. As shown in Fig. \ref{para1}, the PCC values do not perform well when $r=7$. It is evident that deep learning methods can not extract robust features in such small image patches. On the Ottawa dataset, the proposed SAFNet obtains the best performance when $r=9$. When $r>9$, change detection results tend to deteriorate. On the San Francisco and the Yellow River I datasets, the SAFNet achieves the best PCC values when $r=13$. When $r>13$, PCC values tend to get worse. It indicates that it is difficult to describe the changed information of the center pixel by large image patches. Therefore, in our implementation, we set $r=9$ and $r=11$ on the Ottawa and Yellow River II datasets. On the San Francisco and Yellow River I datasets, $r=13$ is employed as the best choice.

\subsubsection{Analysis of the Number of Training Samples}

\begin{figure}[ht]
	\begin{center}
		\includegraphics [width=3.4in]{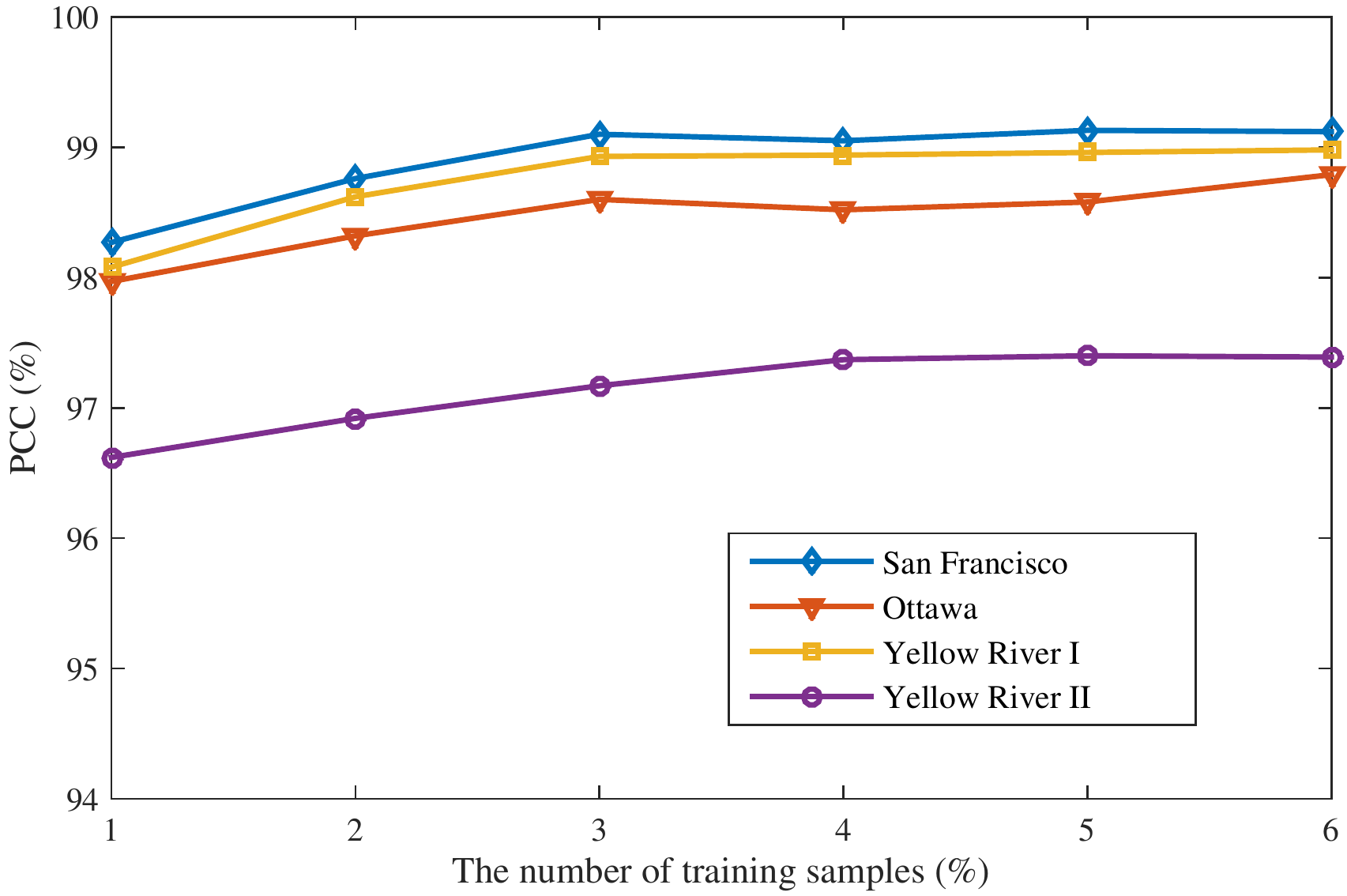}
		\caption{Relationship between the number of training samples $N_t$ and PCC values.}
		\label{para2}
	\end{center}
\end{figure}

Next, we discuss the relationship between the number of training samples $N_t$ and the experimental results, because the training samples are essential. Training samples are randomly selected from pseudo label set. $N_t=[1\%, 2\%, 3\%, 4\%, 5\%, 6\%]$. From Fig. \ref{para2}, we can observe that it is difficult to obtain excellent classification performance with fewer training samples. Because insufficient training samples lead to over-fitting. PCC values commonly decrease with the reduction of $N_t$, especially when $N_t < 3\%$ has a curve of obvious decline. When $N_t > 3\%$, PCC value tends to be stable. Considering that more training samples will affect the efficiency and generalization to a certain extent. Therefore, we choose $N_t=4\%$ on the Yellow River dataset, and $N_t = 4\%$ on the San Francisco and Ottawa datasets.

\subsubsection{Analysis of the Hyperparameter}

In this paper, the combination of similarity measure and classification loss is used to jointly optimize the model parameters. We use a hyperparameter to balance the similarity measure and classification prediction. Here, $\lambda$ is set to $[0.01, 0.1, 0.5, 1]$. It is found from Table \ref{para4} that when $\lambda=1$, it is difficult to focus on the model prediction, which is utilized for the measurement of final results. When $\lambda>0.5$, the PCC values began to decrease gradually, due to the similarity measurement is reduced. It reflects that similarity measure improves discrimination to a certain extent. Therefore, $\lambda=0.5$ is selected as the most appropriate hyperparameter.

\renewcommand\arraystretch{1.4}
\begin{table}[ht]
	\centering 
	\caption{Relationship between PCC and the regularization \\ factor of loss functions.}
	\begin{tabular}{m{1.2cm} | m{1.2cm}<{\centering} m{1.2cm}<{\centering} m{1.2cm}<{\centering} m{1.2cm}<{\centering}}
		\hline
		Dataset & San Francisco   & Ottawa &	Yellow River I &Yellow River II \\ \hline
	$\lambda$=1 & 98.21 & 97.68  & 96.85 & 96.74\\ 
	$\lambda$=0.5 & \textbf{99.10} & \textbf{98.60} & \textbf{98.94} & \textbf{97.37}\\ 
	$\lambda$=0.1 & 98.69 & 98.26 & 98.49 & 96.91 \\ 
	$\lambda$=0.01 & 98.13 & 97.95 & 97.79 & 96.10 \\ 
	\hline
	\end{tabular}
	\label{para4}
\end{table}

\subsubsection{Ablation Experiment}

We present an ablation experiment in which importance of model components is evaluated. Table \ref{para3} shows the relationship among AF module, correlation layer, and PCC values. We can observe that the full model (SAFNet with AF module and correlation layer) achieves better performance. Specifically, there are 0.09\%, 0.58\%,  0.49\%, and 0.33\% improvements in PCC on four datasets compared with the full model excluding AF module, respectively. That is because multiscale responses from convolutional layers are adaptively combined by AF module. Therefore, it is shown that the AF module effectively alleviates the problem of unstable feature representation. In addition, an ablation experiment between the correlation layer and PCC values is exhibited. Table \ref{para3} reveals that the PCC value of the full model is higher than the full model without the correlation layer. That is because the features from two-branch networks further integrate for classification. To sum up, the proposed SAFNet benefits from the AF module and correlation analysis.

\renewcommand\arraystretch{1.4}
\begin{table}[ht]
	\centering 
	\caption{Relationship among AF module, correlation layer \\ and PCC values.}
	\begin{tabular}{m{2.6cm} | m{1.2cm}<{\centering} m{0.8cm}<{\centering} m{0.8cm}<{\centering} m{0.9cm}<{\centering}}
		\hline
		Method & San Francisco   & Ottawa & 
		Yellow River I &Yellow River II \\ \hline
		Full model & \textbf{99.10} & \textbf{98.60} & \textbf{98.94} & \textbf{97.37} \\ 
		w/o AF & 99.01 & 98.02  & 98.45 & 97.04\\ 
		w/o correlation & 98.89 & 98.24 & 98.61 & 97.22\\ 
	    w/o AF and correlation & 98.80 & 97.96 & 98.19 & 96.70 \\ \hline
	\end{tabular}
	\label{para3}
\end{table}

\subsection{Experimental Results and Discussion}

To verify the effectiveness of the proposed SAFNet, several existing change detection methods are used for comparison, i.e., PCAKM \cite{Celik09_grsl}, NBRELM \cite{Gao16_jars}, GaborPCANet \cite{Gao16_grsl}, RMG-FDA \cite{Gao18_jars}, ResNet \cite{He16_cvpr}, DBN \cite{Gong15_tnnls}, DCNet \cite{Gao20_jstars}, and ESCNet \cite{Zhang21_tnnls}. PCAKM uses PCA filters for feature extraction, and then the extracted features are classified by $k$-means clustering. In NBRELM, the neighborhood-based ratio operator is adopted on DI generation and feature extraction. The extracted features are then classified by the extreme learning machine (ELM). GaborPCANet utilizes PCANet to extract discriminant features. PCANet is a simple deep learning network whose convolution filters are chosen from PCA filters. In RMG-FDA, the graph-based method is chosen as the classifier for changed pixel identification. In ResNet, ResNet-18 is employed for pixel-wise classification. DBN applies a deep belief network to complete the SAR change detection task. DCNet is a deep neural network, which cascades multiple channel-weighting based residual blocks. In ESCNet, two weight-share superpixel sampling networks and a siamese neural network based on U-NET are utilized to mine the information between multitemporal images. Specifically, for the PCAKM, the size of neighborhood patch is set to be $S=3$, and the number of principal components selected by PCA is set as $h=5$. In GaborPCANet, $f=\sqrt{2}$, $V=5$, $U=8$, and $k_{max}=2\pi$ are utilized for feature extraction with PCANet. For the NBRELM and GaborPCANet, the size of neighborhood patch is set to be $7\times7$. In RMG-FDA, the threshold for frequency-domain analysis is set as $SmoothVal = 0.1$ and $t_{sal} = 0.3$, and $6$ graphs are used in the random multi-graphs algorithm.

\subsubsection{Results on the San Francisco Dataset}

\begin{figure}[t]
	\begin{center}
		\includegraphics [width=3.4in]{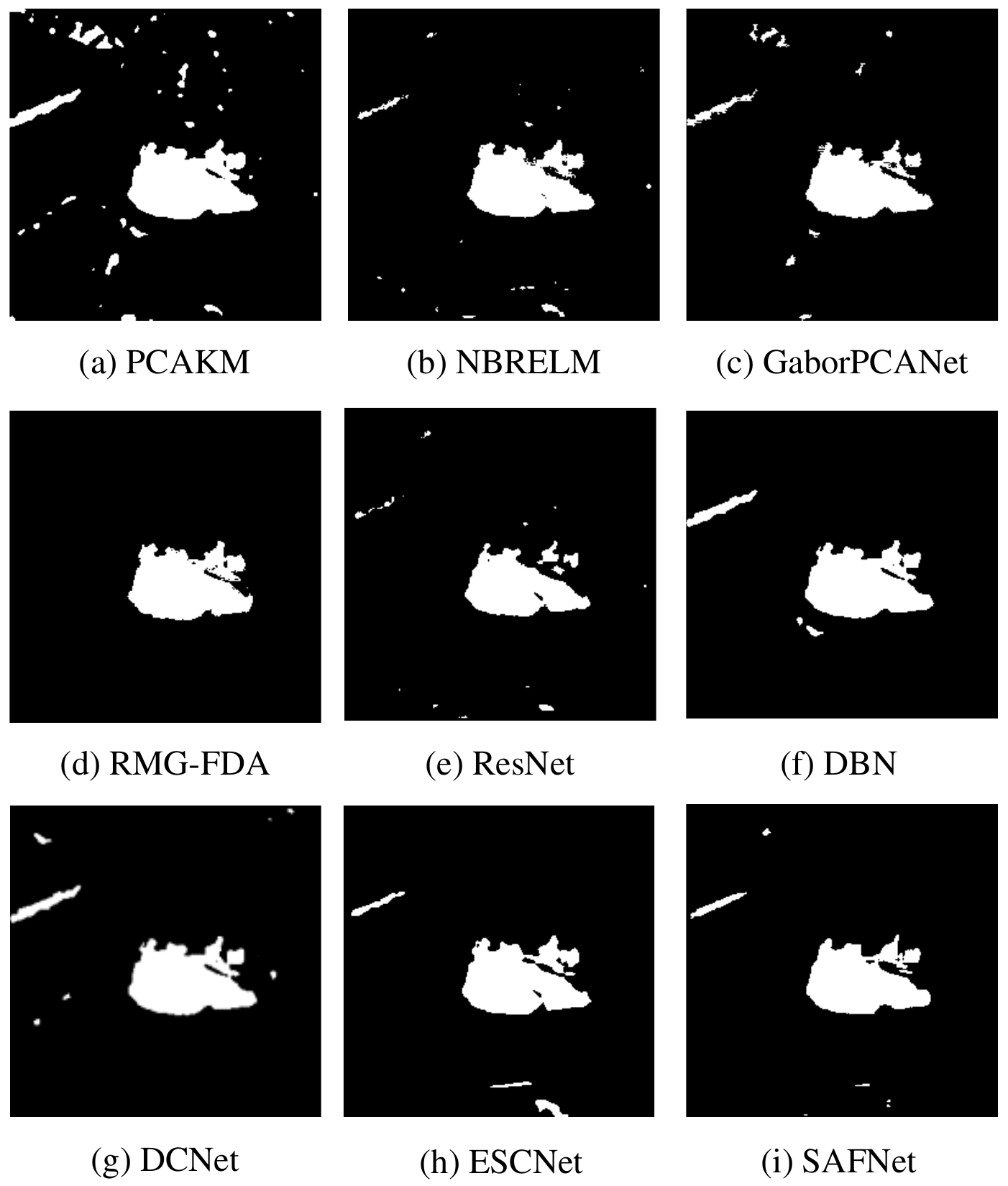}
		\caption{Visualized results of change detection methods on the San Francisco dataset. (a) Result by PCAKM. (b)Result by NBRELM. (c) Result by GaborPCANet. (d) Result by RMG-FDA. (e) Result by ResNet. (f) Result by DBN. (g) Result by DCNet. (h) Result by the proposed ESCNet. (i) Result by the proposed SAFNet.}
		\label{fig_res_san}
	\end{center}
\end{figure}

\renewcommand\arraystretch{1.5}
\begin{table}[h!]
	\centering
	\caption{Change detection results of different methods on the \\ San Francisco dataset.}
	\begin{tabular}{l|ccccc} \hline
		Method & FP & FN & OE & PCC(\%) & KC(\%) \\ \hline
		PCAKM \cite{Celik09_grsl} & 1618 &\textbf{25} & 1643 & 97.49 &  83.68\\
		NBRELM \cite{Gao16_jars} & 393 & 362 & 755 & 98.85 & 91.35\\
		GaborPCANet \cite{Gao16_grsl} & 333 & 342 & 675 & 98.97 & 92.23 \\
		RMG-FDA \cite{Gao18_jars}  &267 &  428 &  695 & 98.94 &91.88 \\
		ResNet \cite{He16_cvpr}  &\textbf{104} &  938 &  1042 & 98.41 & 86.95 \\
		DBN \cite{Gong15_tnnls}  &204&  478 &  682 & 98.96 &91.94 \\
		DCNet \cite{Gao20_jstars}  & 276 &  468 & 744 & 98.86 & 91.28 \\
		ESCNet \cite{Zhang21_tnnls}  &554 & 290 & 844 & 98.71 & 90.54 \\
		Proposed SAFNet & 212 & 377 & \textbf{589} & \textbf{99.10} & \textbf{93.12}\\
		\hline
	\end{tabular}
	\label{table_res_san}
\end{table}

Fig. \ref{fig_res_san} provides the visual comparison of the change maps generated by different methods on the San Francisco dataset. The quantitative evaluation is given in Table \ref{table_res_san}. It can be observed that in the change maps generated by PCAKM, many unchanged pixels are falsely detected as changed ones. Therefore, the FP value of PCAKM is relatively high. RMG-FDA and ResNet achieve better performance with less noise, but many changed pixels are missed. For NBRELM and GaborPCANet, balanced FP and FN values are obtained. However, there is still some white noise in the change maps, and some changed regions are ignored in the upper left corner. More acceptable visualized results are shown in Fig. \ref{fig_res_san} (f)-(i). Compared with the Fig.\ref{fig_res_san} (f) and (g), DCNet obtains the details of the change regions, while DBN suppresses the noise. Furthermore, the proposed SAFNet achieves the best PCC and KC values. It is evident that the AF module and correlation analysis improve the performance of change detection.

\subsubsection{Results on the Ottawa Dataset}

\begin{figure}[t!]
	\begin{center}
		\includegraphics [width=3.4in]{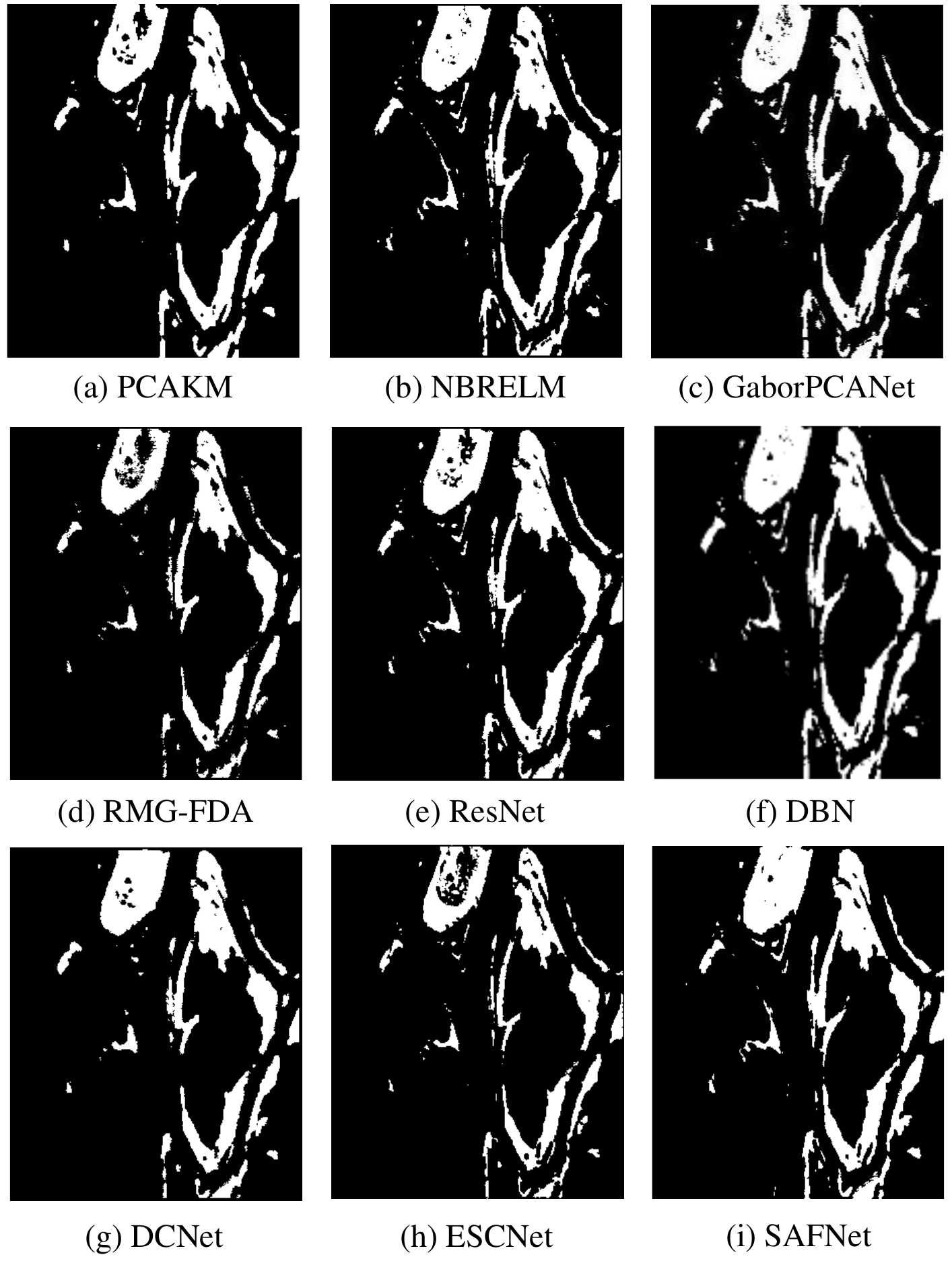}
		\caption{Visualized results of change detection methods on the Ottawa dataset. (a) Result by PCAKM. (b)Result by NBRELM. (c) Result by GaborPCANet. (d) Result by RMG-FDA. (e) Result by ResNet. (f) Result by DBN. (g) Result by DCNet. (h) Result by the proposed ESCNet. (i) Result by the proposed SAFNet.}
		\label{fig_res_ottawa}
	\end{center}
\end{figure}

\renewcommand\arraystretch{1.5}
\begin{table}[h!]
	\centering
	\caption{Change detection results of different methods on the \\ Ottawa dataset.}
	\begin{tabular}{l|ccccc} \hline
		Method & FP & FN & OE & PCC(\%) & KC(\%)\\ \hline
		PCAKM \cite{Celik09_grsl} & 955 & 1515 & 2470 & 97.57 &90.73\\
		NBRELM \cite{Gao16_jars} & 963 & 1192 & 2155 & 97.88 &91.98 \\
		GaborPCANet \cite{Gao16_grsl} & 953 & 942 & 1895 & 98.13 & 92.99 \\
		RMG-FDA \cite{Gao18_jars}  &198 & 1883  & 2081 & 97.95 & 91.96\\
		ResNet \cite{He16_cvpr}  &551 & 1623  & 2174 & 97.86 & 91.73\\
		DBN \cite{Gong15_tnnls}  &995 & 704  & 1699 & 98.33 & 93.76\\
		DCNet \cite{Gao20_jstars}   & 679 & 1051 & 1730 & 98.30 & 93.54\\
		ESCNet \cite{Zhang21_tnnls}  &\textbf{10} & 2231 & 2241 & 97.79 & 86.06 \\
		Proposed SAFNet & 882 & \textbf{534} & \textbf{1416} & \textbf{98.60} & \textbf{94.81}\\
		\hline
	\end{tabular}
	\label{table_res_ottawa}
\end{table}

Fig.\ref{fig_res_ottawa} illustrates the change maps by different methods on the Ottawa dataset. Table \ref{table_res_ottawa} lists the detailed evaluation criteria. It can be observed that the PCAKM and RMG-FDA missed many small changed regions. Therefore, these methods suffer from high FN values. NBRELM, GaborPCANet, and the proposed SAFNet perform better. From Table \ref{table_res_ottawa}, we can observe that the SAFNet is superior to NBRELM, and is about 0.72\% higher in PCC values. Meanwhile, the SAFNet surpasses GaborPCANet by 0.47\% in PCC. In ECANet's results, FP value is the lowest while FN value is high, because many subtle change regions are eliminated as noise. In addition, the PCC of SAFNet is slightly higher than DBN and DCNet. That is because SAFNet has better detail retention resulting in lower FN values. In general, the proposed SAFNet is predominant on the Ottawa dataset compared with some state-of-the-art methods.

\begin{figure}[t!]
	\begin{center}
		\includegraphics [width=3.4in]{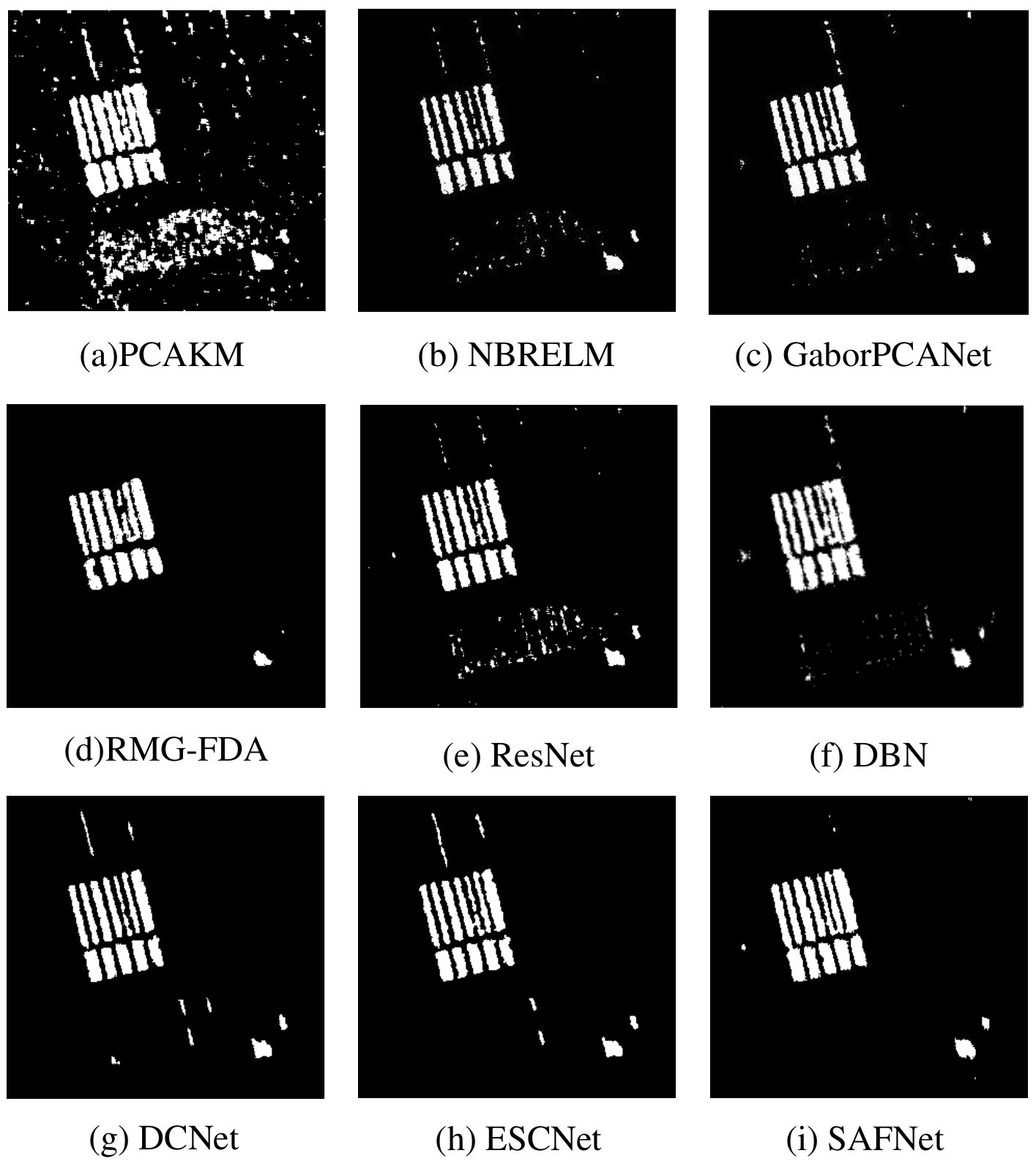}
		\caption{Visualized results of change detection methods on the Yellow River I dataset. (a) Result by PCAKM. (b)Result by NBRELM. (c) Result by GaborPCANet. (d) Result by RMG-FDA. (e) Result by ResNet. (f) Result by DBN. (g) Result by DCNet. (h) Result by the proposed ESCNet. (i) Result by the proposed SAFNet.}
		\label{fig_res_farmland}
	\end{center}
\end{figure}

\renewcommand\arraystretch{1.5}
\begin{table}[h!]
	\centering
	\caption{Change detection results of different methods on the \\ Yellow River I dataset.}
	\begin{tabular}{l|ccccc} \hline
		Method & FP & FN & OE & PCC(\%)& KC(\%)\\ \hline
		PCAKM \cite{Celik09_grsl} & 5158  & \textbf{155} & 5313 & 94.03 & 62.92\\
		NBRELM \cite{Gao16_jars} & 256 & 1794 & 2050 & 97.70 & 76.05\\
		GaborPCANet \cite{Gao16_grsl} & 745 & 1113 & 1858 & 97.91 & 80.63 \\
		RMG-FDA \cite{Gao18_jars} & \textbf{169} & 1614 & 1783 & 98.00 & 79.37\\
		ResNet \cite{He16_cvpr} & 1267 & 457 & 1724 & 98.06 & 83.78 \\
		DBN \cite{Gong15_tnnls} & 561 & 668 & 1229 & 98.62 & 87.49\\
		DCNet \cite{Gao20_jstars}  & 493 & 658 & 1151 & 98.71 & 88.33\\
		ESCNet \cite{Zhang21_tnnls}  & 335 & 916  & 1251 & 98.60 & 86.70\\
		Proposed SAFNet & 359 & 582 &\textbf{941} & \textbf{98.94} &\textbf{90.32} \\
		\hline
	\end{tabular}
	\label{table_res_farmland}
\end{table}

\subsubsection{Results on the Yellow River Dataset}

\begin{figure}[t!]
	\begin{center}
		\includegraphics [width=3.4in]{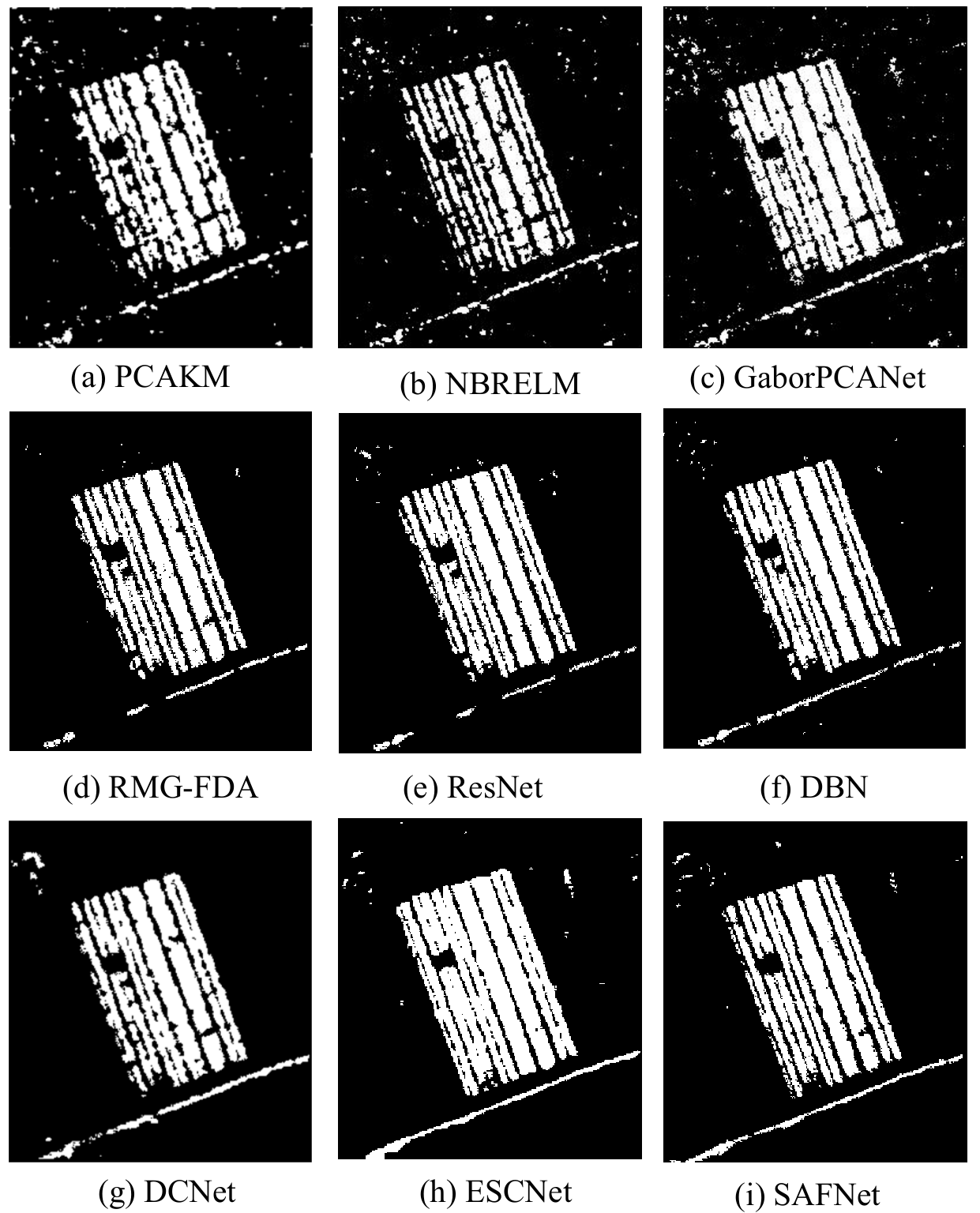}
		\caption{Visualized results of change detection methods on the Yellow River II dataset. (a) Result by PCAKM. (b)Result by NBRELM. (c) Result by GaborPCANet. (d) Result by RMG-FDA. (e) Result by ResNet. (f) Result by DBN. (g) Result by DCNet. (h) Result by the proposed ESCNet. (i) Result by the proposed SAFNet.}
		\label{fig_res_yellow}
	\end{center}
\end{figure}

The change maps generated by different methods on the Yellow River dataset is shown in Fig. \ref{fig_res_farmland} and Fig. \ref{fig_res_yellow}, and the corresponding evaluation criteria are listed in Table \ref{table_res_farmland} and Table \ref{table_res_yellow}. The Yellow River datasets are seriously interfered by speckle noise. Therefore, it is challenging for traditional techniques to obtain satisfying results. For Yellow River I dataset, the change maps of PCAKM and GaborPCANet exhibit many noise regions, and hence both methods suffer from very high FP values. For NBRELM, many changed pixels are missed, and therefore the FN value is relatively high. Despite the noise interference is suppressed, much important change information is ignored and results in a high FN value. Fig. \ref{fig_res_farmland} (f)-(i) perform better, and it is evident that deep learning-based methods can explore contextual information more effectively. Especially, ESCNet and SAFNet suppress the noise at the bottom of the change map.

\renewcommand\arraystretch{1.5}
\begin{table}[h!]
	\centering
	\caption{Change detection results of different methods on the \\ Yellow River II dataset.}
	\begin{tabular}{l|ccccc} \hline
		Method & FP & FN & OE & PCC(\%)& KC(\%)\\ \hline
		PCAKM \cite{Celik09_grsl} & 2137  & 2663 & 4800 & 93.54 & 77.85\\
		NBRELM \cite{Gao16_jars} & 1059 & 3777 & 4836 & 93.49 & 76.14\\
		GaborPCANet \cite{Gao16_grsl} & 2435 & 1533 & 3968 & 94.66 & 82.43  \\
		RMG-FDA \cite{Gao18_jars} & 380 & 1906 & 2286 & 96.92 & 89.13 \\
		ResNet \cite{He16_cvpr} & 1031 & 1870 & 2901 & 96.09 & 86.49 \\
		DBN \cite{Gong15_tnnls} & 861 & 1266 & 2127 & 97.14 & 90.22\\
		DCNet \cite{Gao20_jstars} & 934 & 1809 & 2743 & 96.31 & 87.21\\
		ESCNet \cite{Zhang21_tnnls}  & \textbf{181} & 1981 & 2162 & 97.09 & 89.63\\
		Proposed SAFNet & 734 &\textbf{1222} &\textbf{1956} & \textbf{97.37} & \textbf{90.98}\\
		\hline
	\end{tabular}
	\label{table_res_yellow}
\end{table}

For Yellow River II dataset, we can observe a large number of noise regions in Fig. \ref{fig_res_yellow} (a)-(c). In addition, many changed information is missed at the bottom of changed regions. As a result, the change detection results are seriously deteriorated. For RMG-FDA, noise is effectively suppressed by frequency-domain analysis, but much changed change information is lost. In contrast, Fig. \ref{fig_res_yellow} (f)-(h) generated by deep learning-based methods are more similar to ground truth image. However, the changed information is ignored on the left of the change map resulted by DCNet. Therefore, the FN values are relatively high. Generally speaking, the proposed SAFNet achieves the best performance of change detection. The PCC value of SAFNet is 0.23\%, 1.06\%, and 0.28\% ahead of DBN, DCNet, and ESCNet. According to the visualized result and PCC, SAFNet exhibits the best performance in Yellow River datasets.

Based on the above experiments on four real SAR datasets, the proposed SAFNet has superior performance over several traditional shallow classification models. Besides, by employing the dual network with similarity loss and classification loss to extract the high-level semantic features, the proposed SAFNet achieves better performance than other deep learning-based methods on four real SAR datasets. Furthermore, AF module improves the change detection performance by adaptively fusing the multi-level features for classification. Moreover, the proposed SAFNet has a strong capacity in feature learning. It is a powerful and useful tool for SAR image change detection.

\begin{figure*}[t!]
	\begin{center}
		\includegraphics [width=6.8in]{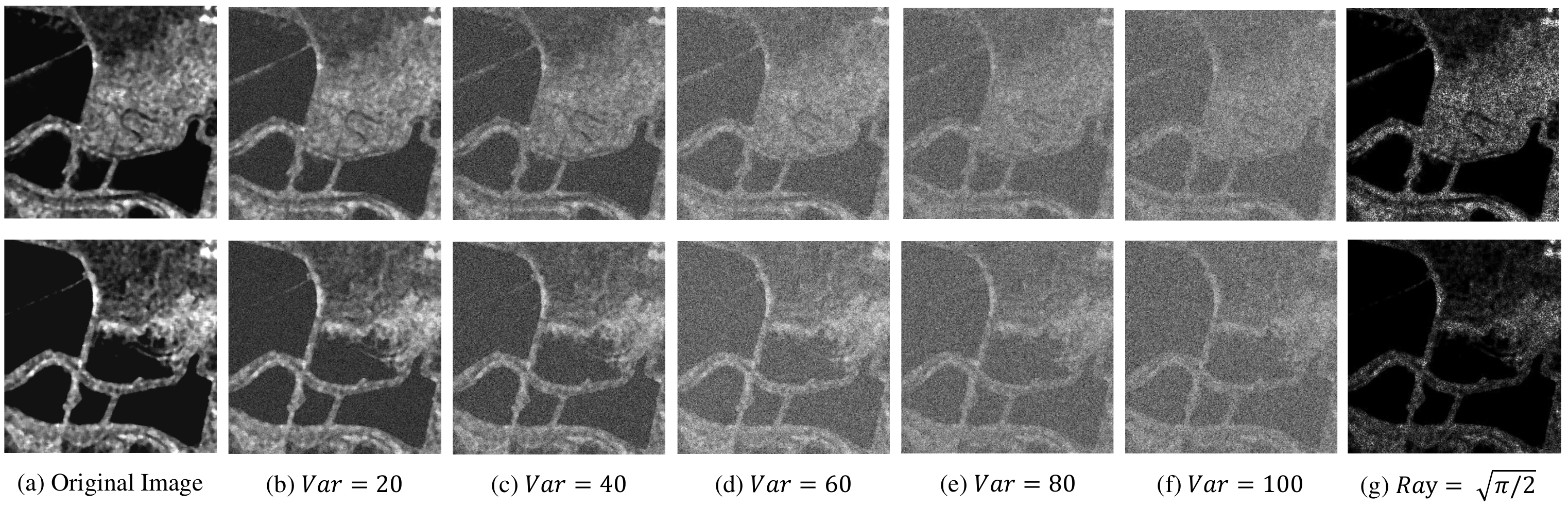}
		\caption{The noise-polluted San Francisco dataset.}
		\label{fig_noise}
	\end{center}
\end{figure*}

\renewcommand\arraystretch{1.8}
\begin{table*}[t]
	\centering 
	\caption{Change detection results on noise polluted datasets.}
	\begin{tabular}{ l |c c c c c c c}
		\hline
		\multirow{2}{*}{Dataset} &
		\multicolumn{7}{c}{PCC value on noise-polluted datasets(\%)}\\
		\cline{2-8} 
		&$Var=0$ & $Var=20$ & $Var=40$ & $Var=60$ & $Var=80$ & $Var=100$ & $Ray=\sqrt{\pi/2}$ \\ \hline
		
		San Francisco  & 99.10  &  98.99  & 98.76  &  98.72 & 98.44 & 98.06 & 98.65\\   
		Ottawa    & 98.60 & 98.23   & 97.95   & 97.51  &  97.20 & 96.75 & 97.82\\ 
		Yellow River I &  98.94  & 98.75  & 98.63   & 98.57  & 98.29  & 98.17 &  98.33\\ 
		Yellow River II & 97.37 & 97.10 & 96.37  &95.89  & 94.65 & 93.70 & 96.09\\ 
		\hline
	\end{tabular}
	\label{table_noise}
\end{table*}

\subsection{Noise Robustness Analysis}

Speckle noise degrades the quality of SAR images. It causes the entanglement between noise and signal. Therefore, we further compare and analyze the noise robustness of the proposed method in this experiment. First, speckle noise is added to the original SAR dataset $I_n=I\cdot n$. Here, $I$ is the original SAR image, $n$ is the noise characteristics. As shown in Table \ref{table_noise} and Table \ref{table_noise}, $Var=40$ represents the noise with Gaussian distribution variance of $40$, and so on. $Ray=\sqrt{\pi /2}$ represents Rayleigh distribution noise with parameter $\sqrt{\pi /2}$. Larger variance means image degradation more seriously. The noise-polluted San Francisco dataset is typically shown in Fig. \ref{fig_noise}, Gaussian noise reduces the signal-to-noise ratio (SNR) of the image, and thus affects the visualization. Rayleigh distribution noise aggravates the interference of multiplicative noise which is difficult to filter out.

We can observe from Table \ref{table_noise} that as the noise level increases, the performance of the change detection method deteriorates rapidly. Especially when $Var> 60$, the PCC values drop sharply. In addition, the results become worse under the interference of Rayleigh distribution noise. Although the original images suffer from different levels of noise, the SAFNet still reaches acceptable PCC values in most cases. In conclusion, the performance of SAFNet declined with along a small scope. That is because the stable feature representation is directly obtained from the input images, thus the noise interference is suppressed to some extent.

\section{Conclusions}

In the past few years, many deep learning methods for SAR image change detection have attracted a lot of attention. However, discriminative feature representation still needs improvement. In this paper, features from different levels are adaptively fused by AF module.  Therefore, meaningful features are emphasized and irrelevant ones are suppressed. In addition, the correlation layer is introduced to further integrate the features from two-branch networks. Besides, feature extraction of two-branch networks is guided by the similarity measure. In the projection space, the changed image pairs still maintain the similarity, while the unchanged pixels are far away from each other. After the correlation layer, the classification loss is employed to optimize the whole network. Therefore, distinctive feature representation is achieved for change detection. Compared with the state-of-the-art methods, the SAFNet exhibits superior performance in terms of quantitative metrics and visual comparison.

With the development of high-resolution Earth observation technology, change detection technology will serve high-resolution, long-time series, and large-scale scenarios in the future. In addition, multisource image change detection is also deserved extensive attention.

\bibliographystyle{IEEEtran}
\bibliography{main.bib}

\begin{thebibliography}{10}
\providecommand{\url}[1]{#1}
\csname url@samestyle\endcsname
\providecommand{\newblock}{\relax}
\providecommand{\bibinfo}[2]{#2}
\providecommand{\BIBentrySTDinterwordspacing}{\spaceskip=0pt\relax}
\providecommand{\BIBentryALTinterwordstretchfactor}{4}
\providecommand{\BIBentryALTinterwordspacing}{\spaceskip=\fontdimen2\font plus
\BIBentryALTinterwordstretchfactor\fontdimen3\font minus
  \fontdimen4\font\relax}
\providecommand{\BIBforeignlanguage}[2]{{%
\expandafter\ifx\csname l@#1\endcsname\relax
\typeout{** WARNING: IEEEtran.bst: No hyphenation pattern has been}%
\typeout{** loaded for the language `#1'. Using the pattern for}%
\typeout{** the default language instead.}%
\else
\language=\csname l@#1\endcsname
\fi
#2}}
\providecommand{\BIBdecl}{\relax}
\BIBdecl

\bibitem{Quin13_tgrs}
G.~Quin, B.~Pinel-Puyssegur, J.-M. Nicolas, and P.~Loreaux, ``Mimosa: An
  automatic change detection method for sar time series,'' \emph{IEEE
  Transactions on Geoscience and Remote Sensing}, vol.~52, no.~9, pp.
  5349--5363, 2013.

\bibitem{Quin10_pr}
X.~Bai and F.~Zhou, ``Analysis of new top-hat transformation and the
  application for infrared dim small target detection,'' \emph{Pattern
  Recognition}, vol.~43, no.~6, pp. 2145--2156, 2010.

\bibitem{Bruzzone97_tgrs}
L.~Bruzzone and S.~B. Serpico, ``An iterative technique for the detection of
  land-cover transitions in multitemporal remote-sensing images,'' \emph{IEEE
  Transactions on Geoscience and Remote Sensing}, vol.~35, no.~4, pp. 858--867,
  1997.

\bibitem{Radke05_tip}
R.~J. Radke, S.~Andra, O.~Al-Kofahi, and B.~Roysam, ``Image change detection
  algorithms: a systematic survey,'' \emph{IEEE Transactions on Image
  Processing}, vol.~14, no.~3, pp. 294--307, 2005.

\bibitem{Jian14_tcb}
M.~Jian, K.-M. Lam, J.~Dong, and L.~Shen, ``Visual-patch-attention-aware
  saliency detection,'' \emph{IEEE Transactions on Cybernetics}, vol.~45,
  no.~8, pp. 1575--1586, 2014.

\bibitem{zhang21_tcb}
M.~Zhang, M.~Gong, H.~He, and S.~Zhu, ``Symmetric all convolutional
  neural-network-based unsupervised feature extraction for hyperspectral images
  classification,'' \emph{IEEE Transactions on Cybernetics}, 2020.

\bibitem{Liu20_tnnls}
J.~Liu, M.~Gong, A.~K. Qin, and K.~C. Tan, ``Bipartite differential neural
  network for unsupervised image change detection,'' \emph{IEEE Transactions on
  Neural Networks and Learning Systems}, vol.~31, no.~3, pp. 876--890, 2020.

\bibitem{Li15_grsl}
H.-C. Li, T.~Celik, N.~Longbotham, and W.~J. Emery, ``Gabor feature based
  unsupervised change detection of multitemporal sar images based on two-level
  clustering,'' \emph{IEEE Geoscience and Remote Sensing Letters}, vol.~12,
  no.~12, pp. 2458--2462, 2015.

\bibitem{Chen19_nc}
H.~Chen, L.~Jiao, M.~Liang, F.~Liu, S.~Yang, and B.~Hou, ``Fast unsupervised
  deep fusion network for change detection of multitemporal sar images,''
  \emph{Neurocomputing}, vol. 332, pp. 56--70, 2019.

\bibitem{Bruzzone02_tip}
L.~Bruzzone and D.~F. Prieto, ``An adaptive semiparametric and context-based
  approach to unsupervised change detection in multitemporal remote-sensing
  images,'' \emph{IEEE Transactions on Image Processing}, vol.~11, no.~4, pp.
  452--466, 2002.

\bibitem{Bazi05_tgrs}
Y.~Bazi, L.~Bruzzone, and F.~Melgani, ``An unsupervised approach based on the
  generalized gaussian model to automatic change detection in multitemporal sar
  images,'' \emph{IEEE Transactions on Geoscience and Remote Sensing}, vol.~43,
  no.~4, pp. 874--887, 2005.

\bibitem{Dekker98_ijrs}
R.~Dekker, ``Speckle filtering in satellite sar change detection imagery,''
  \emph{International Journal of Remote Sensing}, vol.~19, no.~6, pp.
  1133--1146, 1998.

\bibitem{Hou14_jstars}
B.~Hou, Q.~Wei, Y.~Zheng, and S.~Wang, ``Unsupervised change detection in sar
  image based on gauss-log ratio image fusion and compressed projection,''
  \emph{IEEE Journal of Selected Topics in Applied Earth Observations and
  Remote Sensing}, vol.~7, no.~8, pp. 3297--3317, 2014.

\bibitem{Gong12_grsl}
M.~Gong, Y.~Cao, and Q.~Wu, ``A neighborhood-based ratio approach for change
  detection in sar images,'' \emph{IEEE Geoscience and Remote Sensing Letters},
  vol.~9, no.~2, pp. 307--311, 2011.

\bibitem{Krinidis10_tip}
S.~Krinidis and V.~Chatzis, ``A robust fuzzy local information c-means
  clustering algorithm,'' \emph{IEEE Transactions on Image Processing},
  vol.~19, no.~5, pp. 1328--1337, 2010.

\bibitem{Gong12_tip}
M.~Gong, Z.~Zhou, and J.~Ma, ``Change detection in synthetic aperture radar
  images based on image fusion and fuzzy clustering,'' \emph{IEEE Transactions
  on Image Processing}, vol.~21, no.~4, pp. 2141--2151, 2011.

\bibitem{Gao16_grsl}
F.~Gao, J.~Dong, B.~Li, and Q.~Xu, ``Automatic change detection in synthetic
  aperture radar images based on pcanet,'' \emph{IEEE Geoscience and Remote
  Sensing Letters}, vol.~13, no.~12, pp. 1792--1796, 2016.

\bibitem{Gao16_jars}
F.~Gao, J.~Dong, B.~Li, Q.~Xu, and C.~Xie, ``Change detection from synthetic
  aperture radar images based on neighborhood-based ratio and extreme learning
  machine,'' \emph{Journal of Applied Remote Sensing}, vol.~10, no.~4, p.
  046019, 2016.

\bibitem{Wang16_rsl}
S.~Wang, S.~Yang, and L.~Jiao, ``Saliency-guided change detection for sar
  imagery using a semi-supervised laplacian svm,'' \emph{Remote Sensing
  Letters}, vol.~7, no.~11, pp. 1043--1052, 2016.

\bibitem{Gao21_tgrs}
Y.~Gao, W.~Li, M.~Zhang, J.~Wang, W.~Sun, R.~Tao, and Q.~Du, ``Hyperspectral
  and multispectral classification for coastal wetland using depthwise feature
  interaction network,'' \emph{IEEE Transactions on Geoscience and Remote
  Sensing}, pp. 1--15, 2021.

\bibitem{Gong17_tec}
M.~Gong, H.~Li, E.~Luo, J.~Liu, and J.~Liu, ``A multiobjective cooperative
  coevolutionary algorithm for hyperspectral sparse unmixing,'' \emph{IEEE
  Transactions on Evolutionary Computation}, vol.~21, no.~2, pp. 234--248,
  2017.

\bibitem{Jian18_jvcir}
M.~Jian, W.~Zhang, H.~Yu, C.~Cui, X.~Nie, H.~Zhang, and Y.~Yin, ``Saliency
  detection based on directional patches extraction and principal local color
  contrast,'' \emph{Journal of Visual Communication and Image Representation},
  vol.~57, pp. 1--11, 2018.

\bibitem{Li20_tnnls}
Q.~Wang, W.~Huang, Z.~Xiong, and X.~Li, ``Looking closer at the scene:
  Multiscale representation learning for remote sensing image scene
  classification,'' \emph{IEEE Transactions on Neural Networks and Learning
  Systems}, 2020.

\bibitem{Wang21_tgrs}
Q.~Wang, Q.~Li, and X.~Li, ``A fast neighborhood grouping method for
  hyperspectral band selection,'' \emph{IEEE Transactions on Geoscience and
  Remote Sensing}, vol.~59, no.~6, pp. 5028--5039, 2021.

\bibitem{Jian21_esa}
M.~Jian, J.~Wang, H.~Yu, G.~Wang, X.~Meng, L.~Yang, J.~Dong, and Y.~Yin,
  ``Visual saliency detection by integrating spatial position prior of object
  with background cues,'' \emph{Expert Systems with Applications}, vol. 168, p.
  114219, 2021.

\bibitem{Gong15_tnnls}
M.~Gong, J.~Zhao, J.~Liu, Q.~Miao, and L.~Jiao, ``Change detection in synthetic
  aperture radar images based on deep neural networks,'' \emph{IEEE
  Transactions on Neural Networks and Learning Systems}, vol.~27, no.~1, pp.
  125--138, 2015.

\bibitem{Liu16_tnnls}
J.~Liu, M.~Gong, K.~Qin, and P.~Zhang, ``A deep convolutional coupling network
  for change detection based on heterogeneous optical and radar images,''
  \emph{IEEE Transactions on Neural Networks and Learning Systems}, vol.~29,
  no.~3, pp. 545--559, 2016.

\bibitem{Mou18_tgrs}
L.~Mou, L.~Bruzzone, and X.~X. Zhu, ``Learning spectral-spatial-temporal
  features via a recurrent convolutional neural network for change detection in
  multispectral imagery,'' \emph{IEEE Transactions on Geoscience and Remote
  Sensing}, vol.~57, no.~2, pp. 924--935, 2018.

\bibitem{Gao19_grsl}
Y.~Gao, F.~Gao, J.~Dong, and S.~Wang, ``Transferred deep learning for sea ice
  change detection from synthetic-aperture radar images,'' \emph{IEEE
  Geoscience and Remote Sensing Letters}, vol.~16, no.~10, pp. 1655--1659,
  2019.

\bibitem{Wang19_tgrs}
Q.~Wang, Z.~Yuan, Q.~Du, and X.~Li, ``Getnet: A general end-to-end 2-d cnn
  framework for hyperspectral image change detection,'' \emph{IEEE Transactions
  on Geoscience and Remote Sensing}, vol.~57, no.~1, pp. 3--13, 2018.

\bibitem{Liu18_tnnls}
F.~Liu, L.~Jiao, X.~Tang, S.~Yang, W.~Ma, and B.~Hou, ``Local restricted
  convolutional neural network for change detection in polarimetric sar
  images,'' \emph{IEEE Transactions on Neural Networks and Learning Systems},
  vol.~30, no.~3, pp. 818--833, 2018.

\bibitem{Li19_rs}
X.~Li, Z.~Yuan, and Q.~Wang, ``Unsupervised deep noise modeling for
  hyperspectral image change detection,'' \emph{Remote Sensing}, vol.~11,
  no.~3, p. 258, 2019.

\bibitem{Xie17_cvpr}
S.~Xie, R.~Girshick, P.~Doll{\'a}r, Z.~Tu, and K.~He, ``Aggregated residual
  transformations for deep neural networks,'' in \emph{Proceedings of the
  IEEE/CVF Conference on Computer Vision and Pattern Recognition}, 2017, pp.
  1492--1500.

\bibitem{Bengio13_arxiv}
Y.~Bengio, N.~L{\'e}onard, and A.~Courville, ``Estimating or propagating
  gradients through stochastic neurons for conditional computation,''
  \emph{arXiv preprint arXiv:1308.3432}, 2013.

\bibitem{Yang19_nips}
B.~Yang, G.~Bender, Q.~V. Le, and J.~Ngiam, ``Condconv: Conditionally
  parameterized convolutions for efficient inference,'' \emph{arXiv preprint
  arXiv:1904.04971}, 2019.

\bibitem{He16_cvpr}
K.~He, X.~Zhang, S.~Ren, and J.~Sun, ``Deep residual learning for image
  recognition,'' in \emph{Proceedings of the IEEE/CVF Conference on Computer
  Vision and Pattern Recognition}, 2016, pp. 770--778.

\bibitem{Hu18_cvpr}
J.~Hu, L.~Shen, and G.~Sun, ``Squeeze-and-excitation networks,'' in
  \emph{Proceedings of the IEEE Conference on Computer Vision and Pattern
  Recognition}, 2018, pp. 7132--7141.

\bibitem{Li19_cvpr}
X.~Li, W.~Wang, X.~Hu, and J.~Yang, ``Selective kernel networks,'' in
  \emph{Proceedings of the IEEE/CVF Conference on Computer Vision and Pattern
  Recognition}, 2019, pp. 510--519.

\bibitem{Celik09_grsl}
T.~Celik, ``Unsupervised change detection in satellite images using principal
  component analysis and $ k $-means clustering,'' \emph{IEEE Geoscience and
  Remote Sensing Letters}, vol.~6, no.~4, pp. 772--776, 2009.

\bibitem{Gao18_jars}
F.~Gao, X.~Wang, J.~Dong, and S.~Wang, ``Synthetic aperture radar image change
  detection based on frequency-domain analysis and random multigraphs,''
  \emph{Journal of Applied Remote Sensing}, vol.~12, no.~1, p. 016010, 2018.

\bibitem{Gao20_jstars}
Y.~Gao, F.~Gao, J.~Dong, and S.~Wang, ``Change detection from synthetic
  aperture radar images based on channel weighting-based deep cascade
  network,'' \emph{IEEE Journal of Selected Topics in Applied Earth
  Observations and Remote Sensing}, vol.~12, no.~11, pp. 4517--4529, 2019.

\bibitem{Zhang21_tnnls}
H.~Zhang, M.~Lin, G.~Yang, and L.~Zhang, ``Escnet: An end-to-end
  superpixel-enhanced change detection network for very-high-resolution remote
  sensing images,'' \emph{IEEE Transactions on Neural Networks and Learning
  Systems}, 2021.

\end{thebibliography}

\begin{IEEEbiography}[{\includegraphics[width=1in,height=1.25in,clip,keepaspectratio]{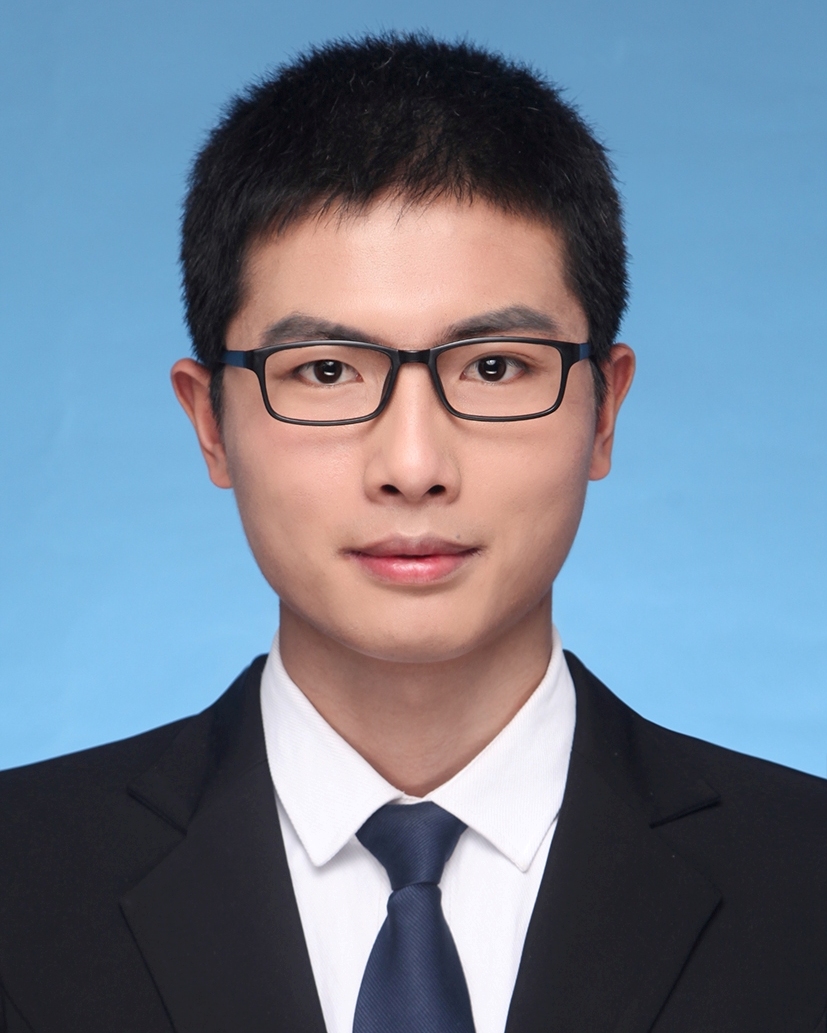}}]{Yunhao Gao} received the B.Sc. degree in electronic information engineering from Ocean University of China, Qingdao, China, in 2017. He is currently pursuing the M.Sc. degree in computer science and applied remote sensing with the School of Information Science and Technology, Ocean University of China, Qingdao, China.

His current research interests include computer vision and remote sensing image processing.

\end{IEEEbiography}

\begin{IEEEbiography}[{\includegraphics[width=1in,height=1.25in,clip,keepaspectratio]{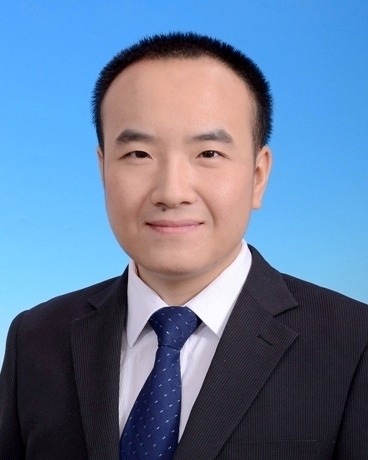}}]{Feng Gao}
received the B.Sc degree in software engineering from Chongqing University, Chongqing, China, in 2008, and the Ph.D. degree in computer science and technology from Beihang University, Beijing, China, in 2015.

He is currently an Associate Professor with the School of Information Science and Engineering, Ocean University of China. His research interests include remote sensing image analysis, pattern recognition and machine learning.

\end{IEEEbiography}

\begin{IEEEbiography}[{\includegraphics[width=1in,height=1.25in,clip,keepaspectratio]{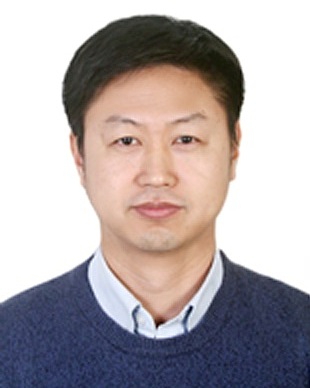}}]{Junyu Dong}
received the B.Sc. and M.Sc. degrees from the Department of Applied Mathematics, Ocean University of China, Qingdao, China, in 1993 and 1999, respectively, and the Ph.D. degree in image processing from the Department of Computer Science, Heriot-Watt University, Edinburgh, United Kingdom, in 2003.

He is currently a Professor and Vice Dean with the School of Information Science and Engineering, Ocean University of China. His research interests include visual information analysis and understanding, machine learning and underwater image processing.
\end{IEEEbiography}

\begin{IEEEbiography}[{\includegraphics[width=1in,height=1.25in,clip,keepaspectratio]{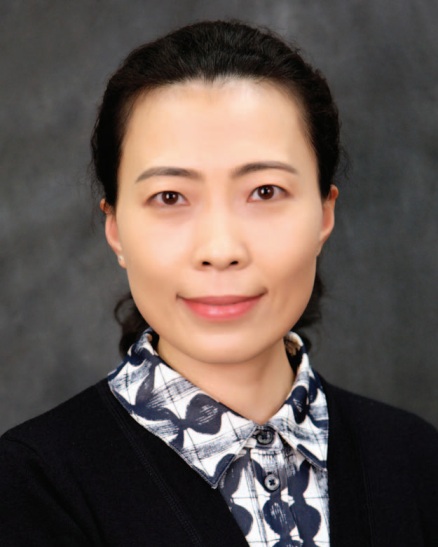}}]{Qian Du}
(Fellow, IEEE) received the Ph.D. degree in electrical engineering from the University of Maryland, Baltimore, MD, USA, in 2000. 

She is currently the Bobby Shackouls Professor with the Department of Electrical and Computer Engineering, Mississippi State University, Starkville, MS, USA. Her research interests include hyperspectral remote sensing image analysis and applications, pattern classification, data compression, and neural networks.

Dr. Du is also a fellow of the SPIE-International Society for Optics and Photonics. She was the Chair of the Remote Sensing and Mapping Technical Committee of the International Association for Pattern Recognition from 2010 to 2014. She was the Co-Chair of the Data Fusion Technical Committee of the IEEE GRSS from 2009 to 2013. She was an Associate Editor of the \textsc{IEEE Journal of Selected Topics in Applied Earth Observations and Remote Sensing} from 2011 to 2015, the \emph{Journal of Applied Remote Sensing} from 2014 to 2015, and the \textsc{IEEE Singal Processing Letters} from 2012 to 2015. She is the Editor-in-Chief of the \textsc{IEEE Journal of Selected Topics in Applied Earth Observations and Remote Sensing} from 2016 to 2020.

\end{IEEEbiography}

\begin{IEEEbiography}[{\includegraphics[width=1in,height=1.25in,clip,keepaspectratio]{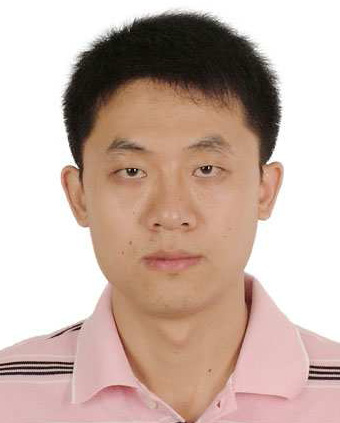}}]{Heng-Chao Li}
(Senior Member, IEEE) received the B.Sc. and M.Sc. degrees from Southwest Jiaotong University, Chengdu, China, in 2001 and 2004, respectively, and the Ph.D. degree from the Graduate University of Chinese Academy of Sciences, Beijing, China, in 2008, all in information and communication engineering. He is currently a Full Professor with the School of Information Science and Technology, Southwest Jiaotong University. His research interests include statistical analysis of synthetic aperture radar (SAR) images, remote sensing image processing, and pattern recognition.

Dr. Li is an Editorial Board Member of the \emph{Journal of Southwest Jiaotong University} and the \emph{Journal of Radars}. He was a recipient of the 2018 Best Reviewer Award from the IEEE Geoscience and Remote Sensing Society for his service to the \textsc{IEEE Journal of Selected Topics in Applied Earth Observations and Remote Sensing} (JSTARS). Moreover, he has served as a Guest Editor for the Special Issues of the \emph{Journal of Real-Time Image Processing}, the IEEE JSTARS, and the IEEE JMASS, a Program Committee Member for the 26th International Joint Conference on Artificial Intelligence (IJCAI-2017), and the 10th International Workshop on the Analysis of Multitemporal Remote Sensing Images (MULTITEMP-2019), and the Session Chair for the 2017 International Geoscience and Remote Sensing Symposium (IGARSS-2017), and the 2019 Asia-Pacific Conference on Synthetic Aperture Radar (APSAR-2019). He is serving as an Associate Editor for the IEEE JSTARS.
\end{IEEEbiography}

\end{document}